\documentclass[aps,prx,showpacs,twocolumn,amsmath,amssymb,nofootinbib]{revtex4-1}
\usepackage{graphicx}
\usepackage{dcolumn}
\usepackage{bm}
\usepackage{amsmath}
\usepackage{amsfonts}
\usepackage{amssymb}
\usepackage{multirow}

\usepackage{amsthm}
\usepackage{pinlabel}
\usepackage{natbib}

\usepackage{epstopdf}
\usepackage[abs]{overpic}
\usepackage[dvipsnames]{xcolor}

\usepackage{tikz}
\usetikzlibrary{arrows.meta}
\usetikzlibrary{decorations.pathmorphing}
\usetikzlibrary{decorations.markings}

\tikzset{
	particle/.style={thick,draw=blue, postaction={decorate},
		decoration={markings,mark=at position .5 with {\arrow[blue]{triangle 45}}}},
	gluon/.style={decorate, draw=black,
		decoration={coil,aspect=0}}
}

\usepackage[dvipsnames]{xcolor}

\usepackage{braket}

\definecolor{specialgray}{HTML}{505050}
\definecolor{col10K}{HTML}{FFA000}
\definecolor{col300K}{HTML}{924FA4}
\definecolor{colMu}{HTML}{5278BD}
\definecolor{colMuI}{HTML}{924FA4}

\definecolor{specialgray}{HTML}{505050}
\definecolor{col10K}{HTML}{FFA000}
\definecolor{col300K}{HTML}{924FA4}
\definecolor{colMu}{HTML}{5278BD}
\definecolor{colMuI}{HTML}{924FA4}
\definecolor{newred}{HTML}{D53E4F}
\definecolor{newblue}{HTML}{5278BD}
\definecolor{newcyan}{HTML}{1EA0A0}
\definecolor{newgreen}{HTML}{5CB14E}
\definecolor{newpurple}{HTML}{924FA4}
\definecolor{newyellow}{HTML}{D1C72E}
\definecolor{neworange}{HTML}{D6923C}

\begin{document}

\title{Multichannel superconductivity of monolayer FeSe on SrTiO$_3$: Interplay of spin fluctuations and electron-phonon interaction} 
\author{Fabian Schrodi}\email{fabian.schrodi@physics.uu.se}
\author{Alex Aperis}\email{alex.aperis@physics.uu.se}
\author{Peter M. Oppeneer}\email{peter.oppeneer@physics.uu.se}
\affiliation{Department of Physics and Astronomy, Uppsala University, P.\ O.\ Box 516, SE-75120 Uppsala, Sweden}

\vskip 0.4cm
\date{\today}

\begin{abstract}
	\noindent 
	We investigate the effects of electron-phonon coupling, as well as of spin and charge fluctuations on the superconducting state in a single layer of FeSe on SrTiO$_3$ substrate. 
	These three bosonic mediators of Cooper pairing are treated on equal footing in a multichannel, full-bandwidth, multiband, and anisotropic Eliashberg theory of the interacting state. Our self-consistent calculations show that an $s$-wave symmetry of the superconducting gap is compatible only with a complete absence of spin fluctuations. When spin fluctuations are present, the sign-changing nodeless $d$-wave pairing symmetry is always obtained, yet the essential ingredient for explaining the gap magnitude and critical temperature is still the interfacial electron-phonon interaction. 
\end{abstract}

\maketitle

Ever since the discovery of superconductivity in monolayer FeSe on SrTiO$_3$ substrate (FeSe/STO)\,\cite{Qing-Yan2012} no consensus has yet been reached about such fundamental questions as, for example, the Cooper pairing mechanism or the superconducting gap symmetry, despite tremendous research effort in both theory and experiment \cite{Kreisel2020}. The measured superconducting critical temperature $T_c$ in FeSe/STO ranges from $50\,\mathrm{K}$ up to over $100\,\mathrm{K}$\,\cite{Qing-Yan2012,Liu2012,He2013,Tan2013,Peng2014,Lee2014,Ge2015}, which is an astonishing increase from the comparatively low $T_c$ ($\sim8\,\mathrm{K}$) of the parent compound FeSe\,\cite{Hsu2008}. The bulk FeSe material has been shown to be nonmagnetic, but is poised in close vicinity to a magnetic phase transition, which is manifested in  strong spin fluctuations (SFs) \cite{Imai2009,Rahn2015,Wang2016NatMat}. It is therefore commonly believed that superconductivity in bulk FeSe, as well as in other Fe-based superconductors with similar Fermi surface (FS) properties, has a magnetic origin\,\cite{Imai2009,Rahn2015,Wang2016NatMat,Wang2016NatCom,Baek2020}.

In FeSe/STO the situation is markedly different because FS nesting conditions are changed due to doping with STO electrons at the interface. The resulting FS consists only of electron-like pockets,
% at the {\red $M$ point of the tetragonal and folded Brillouin zone (BZ)},
 which makes the material inexplicable using `standard' FS nesting arguments. There have, however, been attempts to explain superconductivity in FeSe/STO by theories for SFs that are more specialized to this particular system\,\cite{Linscheid2016a,Kreisel2017}. Recently, experimental \cite{Jandke2019,Liu2019a} and theoretical \cite{Shishidou2018} evidence for magnetic signatures have been provided,
but the literature is sparse on the verification of SFs mediated superconductivity in FeSe/STO. The authors of the current work have argued in Ref.\,\cite{Schrodi2020_3} that SFs possibly contribute to the pairing strength in the superconducting state, but are not the dominant `pairing glue'.

To complicate the issue further, a sizable electron-phonon interaction (EPI) between the substrate phonon and FeSe electrons has been detected in experiment\,\cite{Lee2014,Rebec2017}, an observation reproduced by Density Functional Theory calculations\,\cite{Li2014,Xie2015,Wang2016b,Zhou2016a}. A phonon branch of relatively large frequency $\Omega=81\,\mathrm{meV}$ gives rise to strongly enhanced forward scattering exhibiting small momentum transfer. This characteristic feature of FeSe/STO was shown to play an important role in the superconducting state\,\cite{Rademaker2016,Wang2016c,Aperis2018,Schrodi2018}. However, it is currently debated 
whether EPI and SF effects are competing or cooperative \cite{Song2019}, and how their interplay reflects in experimentally observable quantities of the superconducting state.

This is partially due to uncertainty concerning the symmetry of the superconducting order parameter. Although it has been shown that the gap function does not exhibit nodes in the Brillouin zone (BZ), an observation broadly agreed on, it is still debated whether a sign change occurs between FS pockets separated by a wave vector $\mathbf{q}=(\pi,\pi)$ in the unfolded BZ. In several works it has been argued that the order parameter has $s$-wave symmetry (no sign change), a conclusion drawn from results of impurity measurements \cite{Fan2015} and Angular Resolved Photoemission Spectroscopy\,\cite{Liu2012,Lee2014}. On the contrary, Scanning Tunneling Spectroscopy (STS) carried out in Ref.\,\cite{Liu2018_2} revealed that a sign change of the order parameter ($d$-wave) is similarly possible as $s$-wave, while other recent STS investigations argued more determinedly in favor of a
sign-changing order parameter \cite{Ge2019,LiuPRL2019,Zhang2020}, which suggested SF-mediated pairing to be most relevant \cite{LiuPRL2019}.

In this Letter, we study the competition/cooperation between SFs and EPI in the superconducting state of FeSe/STO. This is done within a self-consistent multichannel Eliashberg formalism in which 
EPI, SFs and charge fluctuations (CFs) are treated on equal footing. 
Our results reveal that the most plausible BZ symmetry of the superconducting gap is nodeless $d$-wave, while an anisotropic $s$-wave state is also possible but only under the exclusion of any influence of spin fluctuations. Hence, our investigation shows that the EPI is responsible for the gap magnitude and high $T_c$, but the SFs generate the unconventional pairing symmetry.

We start from the multiorbital Hubbard-Fr\"ohlich Hamiltonian $\hat{H}=\hat{H}_0+\hat{H}_{\mathrm{int}}+\hat{H}_{\mathrm{ph}}+\hat{H}_{\mathrm{eph}}$, expressed in an orbital basis of the five Fe-$d$ states, with
\begin{align}
&\hat{H}_0=\sum_{\mathbf{k},p,q,\sigma} \xi_{\mathbf{k},p,q} \hat{c}^{\dagger}_{\mathbf{k},p,\sigma}\hat{c}^{~}_{\mathbf{k},q,\sigma}, \\
%\end{align}
%\begin{align}
&\hat{H}_{\mathrm{int}} =  U \sum_{i,s}\hat{n}_{i,s,\uparrow}\hat{n}_{i,s,\downarrow} + \frac{V'}{2} \sum_{i,s,t\neq s} \hat{n}_{i,s}\hat{n}_{i,t} \\\nonumber
& - \frac{J}{2} \sum_{i,s,t\neq s} \hat{\vec{S}}_{i,s}\cdot\hat{\vec{S}}_{i,t}   + \frac{J'}{2} \sum_{i,s,t\neq s,\sigma} \hat{c}^{\dagger}_{i,s,\sigma} \hat{c}^{\dagger}_{i,s,\bar{\sigma}} \hat{c}_{i,t,\bar{\sigma}}\hat{c}_{i,t,\sigma},\\
&\hat{H}_{\mathrm{ph}} =  \hbar\Omega \sum_{\mathbf{q}} \big(\hat{b}_{\mathbf{q}}^{\dagger}\hat{b}^{~}_{\mathbf{q}} + {\textstyle \frac{1}{2}}\big),\\
&\hat{H}_{\mathrm{eph}} =   \sum_{\mathbf{k},\mathbf{k}'}\sum_{p,q,\sigma} g_{\mathbf{q},p,q} \hat{c}^{\dagger}_{\mathbf{k}',p,\sigma}\hat{c}^{~}_{\mathbf{k},q,\sigma}\big(\hat{b}_{\mathbf{q}}^{\dagger}+\hat{b}_{-\mathbf{q}}\big) .
\end{align}	
 Here $\hat{H}_0$ and $\hat{H}_{\mathrm{ph}}$ are the electron and phonon kinetic energies, respectively with $\hat{c}^{\dagger}_{\mathbf{k},p,\sigma}(\hat{b}_{\mathbf{q}}^{\dagger})$
% and $\hat{c}^{~}_{\mathbf{k},p,\sigma}$
 electron (phonon) creation operators, $\sigma$ denotes spin, $p,q,s,t$ orbital indices and $\mathbf{k},\mathbf{k}',\mathbf{q}$ {are} BZ wave vectors (we set $\mathbf{q}=\mathbf{k}-\mathbf{k}'$). 
{We consider the 1-Fe unit cell and {thus} work in the unfolded BZ.} 
 Further, $\xi_{\mathbf{k},p,q}$ denotes electron energies in orbital space and $\Omega$ is a characteristic Einstein phonon frequency. The EPI is given by $\hat{H}_{\mathrm{eph}}$ with EPI scattering matrix elements $g_{\mathbf{q},p,q}$.
  Electron correlations are described by the purely electronic term $\hat{H}_{\mathrm{int}}$ which {carries the information about CFs and SFs mediated interactions}. As usual,  $\hat{\vec{S}}_{i,s} (\hat{n}_{i,s})$ are spin (density) operators, $U$, $V'$ are the respective intra- and inter-orbital Hubbard interactions, $J$ is the Hund's rule coupling, {and} $J'$ the pair-hopping interaction {with} $V'=U-3J/4-J'$ and $J'=J/2$\,\cite{Kubo2007,Graser2009}.
 
Our tight-binding description of FeSe/STO is adopted from Refs.\,\cite{Hao2014,Aperis2018}, where hopping energies for bulk FeSe \cite{Eschrig2009} are modified so as to account for the lattice distortion that arises when a monolayer of FeSe is deposited on the substrate. 
Diagonalization of $\hat{H}_0$ yields electron energies $\xi_{\mathbf{k},n}$ with band index $n$ and matrix elements $a_{\mathbf{k},n}^p$ {which we utilize to derive our Eliashberg theory in band space.} 
%{\red Any chemical potential shift is absorbed in the electron dispersion.}

%%%%%%%%%%%%%%%%%%%%%%%%%%%%%%%%%%
Including the infinite series of Feynman diagrams for all first-order scattering processes due to EPI, SFs, and CFs, we arrive at the electron self-energy\,\cite{Lenck1994},
\begin{align} 
\hat{\Sigma}_{\mathbf{k},m,n} = T \sum_{\mathbf{k}',m',n'} V^{\mathrm{eph}}_{\mathbf{k}-\mathbf{k}',m-m',n,n'} \hat{\rho}_3 \hat{G}_{\mathbf{k}',m',n'} \hat{\rho}_3 \nonumber \\
+ T \sum_{\mathbf{k}',m',n'} V^{\mathrm{S}}_{\mathbf{k}-\mathbf{k}',m-m',n,n'} \hat{\rho}_0 \hat{G}_{\mathbf{k}',m',n'} \hat{\rho}_0 \nonumber \\
+ T \sum_{\mathbf{k}',m',n'} V^{\mathrm{C}}_{\mathbf{k}-\mathbf{k}',m-m',n,n'} \hat{\rho}_3 \hat{G}_{\mathbf{k}',m',n'} \hat{\rho}_3 ,\label{sigma}
\end{align} 
where $m,m'$ index Matsubara frequencies ($\omega_m=\pi T(2m+1)$) and $\hat{\rho}_{0(3)}$ are Pauli matrices.
%
%where $V^{\mathrm{eph}}_{\mathbf{k}-\mathbf{k}',m-m',n,n'}$, $P^{\mathrm{S}}_{\mathbf{k}-\mathbf{k}',m-m',n,n'}$ and $P^{\mathrm{C}}_{\mathbf{k}-\mathbf{k}',m-m',n,n'}$ model the interactions due to phonons, SFs, and CFs, respectively and 
%. The EPI is given by Eq.\,(\ref{epi}), while the spin and charge terms are related to Eq.\,(\ref{vpm}) as
%\begin{align}
%$P^{\mathrm{S}}_{\mathbf{k}-\mathbf{k}',m-m',n,n'} + P^{\mathrm{C}}_{\mathbf{k}-\mathbf{k}',m-m',n,n'} =  V^{(+)}_{\mathbf{q},l,n,n'}$, $P^{\mathrm{S}}_{\mathbf{k}-\mathbf{k}',m-m',n,n'} - P^{\mathrm{C}}_{\mathbf{k}-\mathbf{k}',m-m',n,n'} =  V^{(-)}_{\mathbf{q},l,n,n'}$ .
%\end{align}
%%%%%%%%%%%%%%%%%%%%%%%%%%%%%%%%%%%%%%%%%%%%%%%%
%
The EPI kernel $V^{(\mathrm{eph})}_{\mathbf{q},n,n'}(i\omega_m-i\omega_{m'})$ is derived similarly as in Refs.\,\cite{Aperis2018,Schrodi2020_3}, where the electron-phonon scattering at the interface is  modeled by the functional form $g_{\mathbf{q}} = g_0\exp\big(-|\mathbf{q}|/q_c\big)$ with interaction strength $g_0$,
 $q_c=0.3a^{-1}$ and $a$ the lattice constant
% of FeSe
 \,\cite{Lee2014}. Further, we employ an Einstein phonon with frequency $\Omega=81\,\mathrm{meV}$ to which the FeSe electrons are coupled\,\cite{Lee2014,Xie2015}.  For brevity we use henceforth the notation $V^{(\mathrm{eph})}_{\mathbf{q},l,n,n'}$ with $l=m-m'$. As is described in detail in the {Supplementary Material (SM)} 
  and Ref.\,\cite{Schrodi2020_3}, we keep the full orbital content encoded in $a_{\mathbf{k},n}^p$ when calculating band-dependent interaction kernels $V^{(\pm)}_{\mathbf{q},l,n,n'}=V^{(\mathrm{S})}_{\mathbf{q},l,n,n'}\pm V^{(\mathrm{C})}_{\mathbf{q},l,n,n'}$ for SFs (S) and CFs (C). Labels $(+)$ and $(-)$ are respectively referring to kernels as they are used in electron-energy renormalization and superconducting {equations}, see below.

{From the Green's function $\hat{G}^{-1}_{\mathbf{k},m,n} = i\omega_m Z_{\mathbf{k},m,n}\hat{\rho}_0 - \big(\xi_{\mathbf{k},n} + \Gamma_{\mathbf{k},m,n}\big)\hat{\rho}_3 - \phi_{\mathbf{k},m,n}\hat{\rho}_1$,} we derive a self-consistent set of {a total of 15} {coupled} Eliashberg equations for the mass $Z_{\mathbf{k},n,m}$ and chemical potential $\Gamma_{\mathbf{k},n,m}$ renormalization functions, and the superconducting order parameters $\phi_{\mathbf{k},n,m}$\,\cite{Migdal1958,Eliashberg1960}. 
%These functions carry electronic properties and therefore depend on {\blue electron} Matsubara frequencies $\omega_m=\pi T(2m+1)$, $m\in\mathbb{Z}$. 
The full interaction kernels for EPI, CFs, and SFs are given by $K^{(\pm)}_{\mathbf{q},l,n,n'}=V^{(\mathrm{eph})}_{\mathbf{q},l,n,n'}\pm V^{(\pm)}_{\mathbf{q},l,n,n'}$, using $K^{(+)}_{\mathbf{q},l,n,n'}$ in the equations for $Z_{\mathbf{k},n,m}$ and $\Gamma_{\mathbf{k},n,m}$, and $K^{(-)}_{\mathbf{q},l,n,n'}$ in the  equation for $\phi_{\mathbf{k},n,m}$. For further details we refer to the SM 
%\ref{appTheory}
 and Refs.\,\cite{Schrodi2020_3,Aperis2018}.

In the theory employed here we treat the scattering strength $g_0$ as 
a parameter to control the strength of the EPI. For CFs and SFs we are free to choose the intraorbital onsite interaction $U$ and the Hund's rule coupling $J$.
 For convenience we set $J$ to a fixed ratio of $U$, leaving us with two variational quantities $g_0$ and $U$. Hence, we have direct control over the interaction strength for all three mediators of superconductivity. Our calculations are carried out using the Uppsala Superconductivity (UppSC) code\,\cite{UppSC,Aperis2015,Bekaert2018,Schrodi2019,Schrodi2020_2}, in particular, by combining the advances of Refs.\,\cite{Aperis2018} and \cite{Schrodi2020_3}.

\begin{figure}[t!]
	\centering
	\includegraphics[width=1\linewidth]{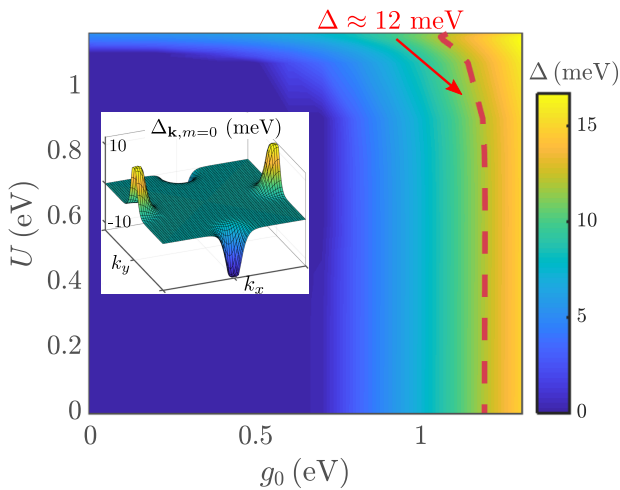}
	\vspace*{-0.3cm}
	\caption{Self-consistently calculated maximum value of the superconducting gap as function of $g_0$ and $U$. The red dashed line corresponds to a value $\Delta\simeq12\,\mathrm{meV}$, compatible with experiment \cite{Lee2014,Liu2012}. {Inset: {the computed} $d$-wave gap function {shown} in the BZ.}}\label{gap}
\end{figure}

Motivated by our earlier work \cite{Schrodi2020_3}, we choose the temperature  $T=5\,\mathrm{K}$ (to compare to available experiments \cite{Lee2014,Liu2012}) and select $J=U/2$. 
{For calculations of the SF and CF kernels we apply a {high-}energy cutoff $\omega_{\mathrm{cut}}=0.54\,\mathrm{eV}$} {(see \cite{Schrodi2020_3})}.
%We restrict the interactions to low energies by a cutoff $\omega_{\mathrm{cut}}=0.54\,\mathrm{eV}$ for calculations of the SF and CF kernels.
 We consider this pair of $(J,\omega_{\mathrm{cut}})$ as it reportedly allows for a finite superconducting gap, even in the absence of any EPI\,\cite{Schrodi2020_3}. Left with two 
%free 
parameters $U$ and $g_0$, which respectively control the strength of CFs/SFs and EPI, we solve the full bandwidth, multiband, and anisotropic Eliashberg equations (see {SM})
% \ref{appTheory}
for each pair $(U,g_0)$. Computing the experimentally observable gap function as $\Delta_{\mathbf{k},m,n}=\phi_{\mathbf{k},m,n}/Z_{\mathbf{k},m,n}$, we plot the maximum value $\Delta=\mathrm{max}_{\mathbf{k},n}\,|\Delta_{\mathbf{k},m=0,n}|$ in Fig.\,\ref{gap}. The red dashed line through $(U,g_0)$-space represents a gap size of $\Delta\simeq12\,\mathrm{meV}$ as was measured for this material\,\cite{Zhang2016,Liu2012}.
%\cite{Zhang2016,Lee2014,Liu2012,Zhang2016}.

It is apparent that the onset of superconductivity with respect to the electron-phonon scattering strength lies at $g_0\sim
0.6 - 0.7\,\mathrm{eV}$. For growing $g_0$ the maximum superconducting gap increases approximately linear for fixed $U$. On the other hand, for small $g_0$ the choice of $U$ must be close to the maximally allowed value, which in term is dictated by the Stoner criterion, to obtain a finite $\Delta$. Here we do not pay special attention to either of the limiting cases $U=0\,\mathrm{eV}$ or $g_0=0\,\mathrm{eV}$, because these have been analyzed in detail in  previous studies\,\cite{Schrodi2020_3,Aperis2018,Schrodi2018}. 

When computing the results of Fig.\,\ref{gap} we assumed that the symmetry of the order parameter is \textit{a priori} not known, which is why we performed each calculation with an initial $s$-wave and $d$-wave state. 
%{\blue Once} both symmetries are stable/converged, we determine the ground state of the system by computing the free energy difference
%\begin{align}
%&\Delta F^{(AB)} = -\frac{T}{2} \sum_{\mathbf{k},m,n} \mathrm{Tr} \big[ \log\big( (\hat{G}_{\mathbf{k},m,n}^A)^{-1}/(\hat{G}_{\mathbf{k},m,n}^B)^{-1} \big) \big] \nonumber \\
%&~~~~~~~ - \frac{T}{4} \sum_{\mathbf{k},m,n} \mathrm{Tr}\big[ \hat{\Sigma}_{\mathbf{k},m,n}^A \hat{G}_{\mathbf{k},m,n}^A - \hat{\Sigma}_{\mathbf{k},m,n}^B \hat{G}_{\mathbf{k},m,n}^B \big] \label{deltaf}
%\end{align}
%between two solutions $(A)$ and $(B)$. Here $\hat{\Sigma}$ and $\hat{G}$ are the electron self-energy and Green's function, see Appendix \ref{appTheory}. The outcome of Eq.\,(\ref{deltaf}) is interpreted as follows:
%\begin{itemize}
%	\item{} $\Delta F^{(AB)}<0$: $A$ is the ground state;
%	\item{} $\Delta F^{(AB)}>0$: $B$ is the ground state;
%	\item{} $\Delta F^{(AB)}=0$: $A$ and $B$ are degenerate.
%\end{itemize}
For all parameter space shown in Fig.\,\ref{gap} we find a $d$-wave symmetry as {the converged solution ({gap shape} shown in the inset of Fig.\,\ref{gap})}, 
with exception of {the purely phononic case ($U=0$)}, 
{where the symmetry is $s$-wave \cite{Aperis2018}}.

\begin{figure}[t!]
	\centering
	\includegraphics[width=1\linewidth]{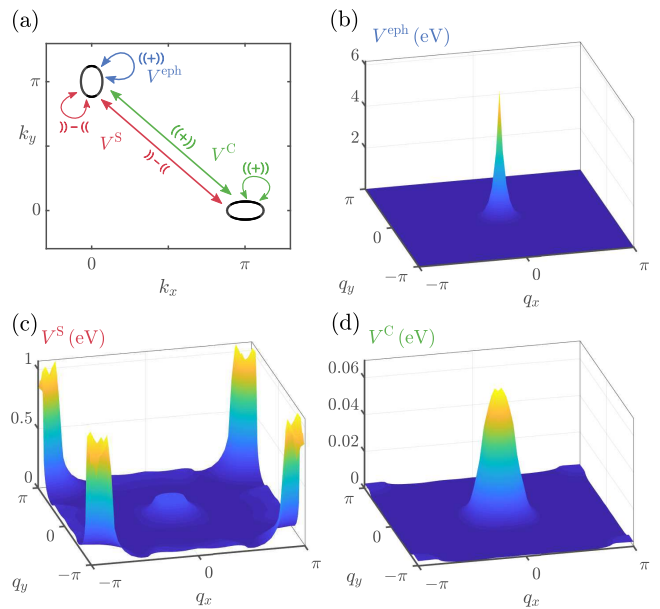}
	\caption{(a) {Multichannel pairing interactions,} 
	%Main interactions in the superconducting channel, 
	schematically drawn on the Fermi surface considered here ({black curve}). Red, blue, and green colors refer, respectively, to spin fluctuations ($V^{\mathrm{S}}$), electron-phonon coupling ($V^{\mathrm{eph}}$), and charge fluctuations ($V^{\mathrm{C}}$). Repulsive interactions are labeled as `$))-(($' {and} attractive couplings as `$((+))$'. (b) Electron-phonon coupling calculated for $g_0=0.5\,\mathrm{eV}$. (c)/(d) Spin/Charge fluctuations kernel for $U=1.07\,\mathrm{eV}$.} \label{kern}
\end{figure}

To better understand these findings let us take a closer look at the couplings in the superconducting channel, where we focus on the FS for simplicity. In Fig.\,\ref{kern}(a) we show the FS sheets of our tight-binding model (black lines) and schematically draw all interactions included in our theory. Here we use the definitions 
%$V^{\mathrm{eph}}=V^{(\mathrm{eph})}_{\mathbf{q},l=0,n\in\mathrm{FS},n'=n}$, $V^{\mathrm{S}}=V^{(\mathrm{S})}_{\mathbf{q},l=0,n\in \mathrm{FS},n'=n}$, and $V^{\mathrm{C}}=V^{(\mathrm{C})}_{\mathbf{q},l=0,n\in\mathrm{FS},n'=n}$ for brevity.
$V^{\mathrm{eph}}= V^{(\mathrm{eph})}_{\mathbf{q},l=0,n,n'}$, $V^{\mathrm{S}}=V^{(\mathrm{S})}_{\mathbf{q},l=0,n,n'}$, and $V^{\mathrm{C}}=V^{(\mathrm{C})}_{\mathbf{q},l=0,n,n'}$, each averaged on the FS, for brevity.
The EPI, shown in Fig.\ \ref{kern}(b) for $g_0=0.5\,\mathrm{eV}$, peaks at small momentum transfer and does not contain any large-$\mathbf{q}$ contributions. Since it enters with a positive sign into the equation of the superconducting gap, no local sign change (on each FS pocket) is promoted. 
Therefore, if 
%we assume that 
no spin and charge fluctuations are taken into account ($U=0$) the  $s$-wave symmetry is favored because  $V^{\mathrm{eph}}$ is 
 attractive. 

Choosing $U=1.07\,\mathrm{eV}$ as an example, we show in Fig.\,\ref{kern}(c) and (d) the SF and CF kernel, respectively. We observe
that a close-to-nesting condition between the two FS pockets leads to a leading contribution at $\mathbf{q}=(\pi,\pi)$ {($=M$)} for the spin part. Further, we find a small-$\mathbf{q}$ coupling for $V^{\mathrm{S}}$ which has significantly lower magnitude. As indicated in panel (a), {since} the SF kernel 
%is repulsive,
{enters repulsively
 in the equation for the superconducting
order parameter (see SM),}
  a sign change {in the superconducting gap} is promoted both globally between the FS sheets, and locally on each pocket. However, the repulsive small-$\mathbf{q}$ contribution of $V^{\mathrm{S}}$ is too weak to induce such a local sign change, 
{which} 
% and it 
 has not been found here {nor} in Ref.\,\cite{Schrodi2020_3}. As concerns CFs, we find a weak attractive interaction peaked at $\mathbf{q}=\Gamma$, see Fig.\,\ref{kern}(d), which has a  similar order of magnitude as the repulsive SFs coupling at this wave vector.

To first order approximation we can assume that SF and CF kernels do not significantly contribute at {$\mathbf{q}=\Gamma$},  
%This is 
due to their comparable magnitude and the fact that $V^{\mathrm{S}}$ and $V^{\mathrm{C}}$ are competing in the superconducting channel. Therefore, we are left with $V^{\mathrm{eph}}$ peaked at $\Gamma$, and $V^{\mathrm{S}}$ having a leading contribution at $M$. In other words, we have a locally sign-conserving EPI, and a repulsive interaction $V^{\mathrm{S}}$ that induces a sign change between FS pockets. 
%Note, that 
These contributions \textit{cooperatively} support a  sign-changing $d$-wave solution. 
%{\red , i.e., the aforementioned degeneracy is broken {\blue when} 
%%due to choosing 
%$U$ {\blue is} finite}.
 Note that for a global $s$-wave state we face a different situation. If the order parameter does not change sign between the two FS pockets, contributions $V^{\mathrm{eph}}$ and $V^{\mathrm{S}}$ are competing and hence reducing the size of the superconducting gap.
 {Then, $s$-wave symmetry becomes energetically less favorable.}
%  {\blue We find that} such a self-consistent solution cannot be stabilized for all pairs $(U,g_0)$ of Fig.\,\ref{gap}, and even if possible, the free energy calculation via Eq.\,(\ref{deltaf}) reveals it not to be the system's ground state.

\begin{figure}[t!]
	\centering
	\includegraphics[width=1\linewidth]{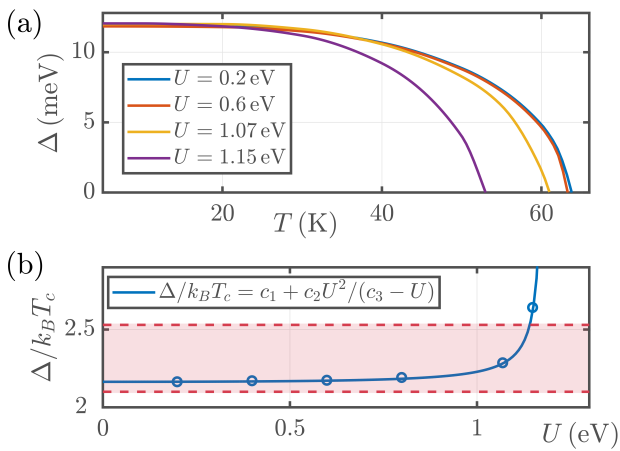}
	\caption{(a) Self-consistently calculated maximum gap as function of temperature for various $U$ as indicated in the legend. (b) Ratio $\Delta/k_BT_c$ as a function of $U$. The reference lines corresponding to $\Delta/k_BT_c = 2.1$ \cite{He2013} and 2.53 \cite{Zhang2016} are plotted in dashed red. Blue circles show our computed data points, which are fitted by the solid blue line, see legend.}\label{deltatc}
\end{figure}

Next, we examine the temperature dependence of the superconducting gap for various pairs of $(U,g_0)$. The couplings are taken from the red dashed line in Fig.\,\ref{gap} such that the magnitude of $\Delta\simeq12\,\mathrm{meV}$ corresponds to the experimental value at $T\sim5\,\mathrm{K}$. In Fig.\,\ref{deltatc}(a) we show the result for the maximum superconducting gap as function of $T$, self-consistently computed for choices of $U$ as written in the legend. 
{Note that} all four curves exhibit the behavior $\underset{T\rightarrow0}{\lim}\Delta(T)\simeq12\,\mathrm{meV}$, while the values for $T_c$ change with $U$. Notably, our results for $U\lesssim0.6\,\mathrm{eV}$, indicating a weak coupling to SFs, fall on top of each other to a good approximation. As $U$ increases towards the maximum value $U_{\mathrm{max}}$, which is dictated by the Stoner criterion\,\cite{Schrodi2020_3}, $T_c$ gradually decreases.

To elucidate this aspect further we draw the ratio $\Delta/k_BT_c$ as blue circles in Fig.\,\ref{deltatc}(b). 
%The experimental reference value $\Delta/k_BT_c \simeq 2.1$ is indicated by the red dashed line\,\cite{He2013}. 
It is apparent that an enhancement of the SF kernel leads to a more strongly coupled system. We analyze this behavior more detailedly by fitting our results to the functional form
\begin{align}
\frac{\Delta}{k_BT_c} = c_1 + c_2 \cdot \frac{U^2}{c_3-U} ~, \label{fitting}
\end{align}
where $c_1$, $c_2$, and $c_3$ are free parameters of the fit. The result, drawn as solid blue line in Fig.\,\ref{deltatc}(b), matches our data points very accurately. 

Taking the limit $U\rightarrow0$ in Eq.\,(\ref{fitting}) corresponds to the case of only considering EPI. Therefore, the value $c_1=2.164$ equals $\Delta/k_BT_c$ without any influence of CFs or SFs, compare Ref.\,\cite{Aperis2018}. Further, it was shown in Ref.\,\cite{Schrodi2020_3} that the leading spin contributions to the kernel at {$\mathbf{q}=M$} scale like $U^2/(U_{\mathrm{max}}-U)$, which makes us associate $c_3\equiv U_{\mathrm{max}}=1.183\,\mathrm{eV}$. This value is remarkably close to the precise value found from the Stoner criterion\,\cite{Schrodi2020_3}. With scaling constant $c_2=0.0118\,\mathrm{eV}^{-1}$ we therefore find that enhanced contributions due to SFs drive $\Delta/k_BT_c$ more towards the strong coupling regime.

The red dashed lines in Fig.\,\ref{deltatc}(b) serve as approximate bounding box for reasonable magnitudes of $\Delta/k_BT_c$, according to existing works on FeSe/STO where the gap magnitude was found as $\sim12\,\mathrm{meV}$\,\cite{Liu2012,Zhang2016,Wang2016c}. For our fitting curve (blue) to stay within the red shaded area, $U$ can be chosen relatively large but not in too close vicinity of the maximally allowed value $U_{\mathrm{max}}$. {This has the consequence that the influence of SFs on the superconducting gap magnitude is bounded.}

Further analysis shows that the decrease of $T_c$ with growing $U$ in Fig.\,\ref{deltatc}(a) stems from a competition of SFs and EPI in {mediating superconductivity.}
%the superconducting pairing.
 To prove this we first need to calculate the renormalized Fermi surface, defined by the condition $\xi_{\mathbf{k},n}+\Gamma_{\mathbf{k},m=0,n}=0$, in the interacting state for a given $T$. No significant changes in comparison to the non-interacting FS are detected in the whole temperature range considered here, which goes in line with earlier predictions of a temperature-independent FS in this system\,\cite{Schrodi2018,Shigekawa2018}. As the influence of SFs increases with $U$, we detect decreasing values of $\langle \phi_{\mathbf{k},m=0,n\in\mathrm{FS}}\rangle_{\mathbf{k}_F}$, while $\lambda_{\mathrm{m}}^{(T<T_c)}=\langle Z_{\mathbf{k},m=0,n\in\mathrm{FS}}\rangle_{\mathbf{k}_F}-1$ grows. Therefore we observe a decrease in the superconducting gap magnitude $\Delta\sim\phi/Z$. 
%Notably, this statement holds true only for the superconducting state, i.e., for $T<T_c$.

We have already seen that SFs and EPI {can} act {\textit{cooperatively}} for 
%the 
%superconducting 
%{\blue Cooper pairing},
{superconductivity,}
%channel, 
a statement that holds true when considering the leading contributions at $M$ and $\Gamma$, respectively. However, an increase in $U$ enhances also the interaction kernel $V^{\mathrm{S}}$ at $\Gamma$, which competes with the small-$\mathbf{q}$ EPI. Therefore, the influence of $V^{\mathrm{eph}}$ gets partially suppressed, which 
%seemingly 
is not compensated  by an increased  $V^{\mathrm{S}}$ coupling at $M$. This is a  {\textit{competing}} aspect of {these} bosonic mediators of {Cooper pairing} {in FeSe/STO.}
% {\blue which} is further backed up by looking into the normal state, setting $T=75\,\mathrm{K}>T_c$. There, the coupling strength $\lambda_{\mathrm{m}}=\langle Z_{\mathbf{k},m=0,n\in\mathrm{FS}}\rangle_{\mathbf{k}_F}-1$ is a constant to {\blue a} good approximation, i.e., not a function of $U$, {\blue in contrast to its $U$ dependence in the superconducting state}. 
 Our computed small-coupling value $\lambda_{\mathrm{m}}\simeq0.35$ is  furthermore consistent with experimental observations\,\cite{Lee2014,Tian2016}.
Importantly, our multichannel calculations show that in the presence of a sizable EPI, even modest SFs  (i.e., small $U$) nonetheless lead to a %sign-changing 
nodeless $d$-wave pairing symmetry. This unconventional symmetry is consistent with the sign-changing order parameter that was deduced in recent STS experiments 
\cite{Zhang2020,LiuPRL2019,Ge2019}.

In summary, we have investigated the superconducting state of FeSe/STO treating EPI, SFs and CFs on equal footing. 
Our self-consistent multichannel Eliashberg theory shows unambiguously that an $s$-wave symmetry of the order parameter is possible only with negligible magnetic contributions, {i.e., in the limit $U\rightarrow0$}. %{\blue On the other hand,} 
{Conversely,} a nodeless
$d$-wave state is realized 
for any small influence of SFs. For the latter
scenario the obtained maximum superconducting gap and $T_c$ are 
compatible with experiment, provided that $U$ is  reasonably
smaller than its maximally possible value (given by the Stoner criterion). Consequently, we are led to identify the main `pairing glue' in FeSe/STO as the EPI,
 but any small contribution from SFs is sufficient to cause an unconventional $d$-wave pairing symmetry.

%\begin{acknowledgments}
	This work has been supported by the Swedish Research Council (VR), the R{\"o}ntgen-{\AA}ngstr{\"o}m Cluster, the K.\ and A.\ Wallenberg Foundation (grant No.\ 2015.0060), and the Swedish National Infrastructure for Computing (SNIC).	
%\end{acknowledgments}

\bibliographystyle{apsrev4-1}
%\bibliography{reportBib.bib}{}

\begin{thebibliography}{50}%
	\makeatletter
	\providecommand \@ifxundefined [1]{%
		\@ifx{#1\undefined}
	}%
	\providecommand \@ifnum [1]{%
		\ifnum #1\expandafter \@firstoftwo
		\else \expandafter \@secondoftwo
		\fi
	}%
	\providecommand \@ifx [1]{%
		\ifx #1\expandafter \@firstoftwo
		\else \expandafter \@secondoftwo
		\fi
	}%
	\providecommand \natexlab [1]{#1}%
	\providecommand \enquote  [1]{``#1''}%
	\providecommand \bibnamefont  [1]{#1}%
	\providecommand \bibfnamefont [1]{#1}%
	\providecommand \citenamefont [1]{#1}%
	\providecommand \href@noop [0]{\@secondoftwo}%
	\providecommand \href [0]{\begingroup \@sanitize@url \@href}%
	\providecommand \@href[1]{\@@startlink{#1}\@@href}%
	\providecommand \@@href[1]{\endgroup#1\@@endlink}%
	\providecommand \@sanitize@url [0]{\catcode `\\12\catcode `\$12\catcode
		`\&12\catcode `\#12\catcode `\^12\catcode `\_12\catcode `\%12\relax}%
	\providecommand \@@startlink[1]{}%
	\providecommand \@@endlink[0]{}%
	\providecommand \url  [0]{\begingroup\@sanitize@url \@url }%
	\providecommand \@url [1]{\endgroup\@href {#1}{\urlprefix }}%
	\providecommand \urlprefix  [0]{URL }%
	\providecommand \Eprint [0]{\href }%
	\providecommand \doibase [0]{http://dx.doi.org/}%
	\providecommand \selectlanguage [0]{\@gobble}%
	\providecommand \bibinfo  [0]{\@secondoftwo}%
	\providecommand \bibfield  [0]{\@secondoftwo}%
	\providecommand \translation [1]{[#1]}%
	\providecommand \BibitemOpen [0]{}%
	\providecommand \bibitemStop [0]{}%
	\providecommand \bibitemNoStop [0]{.\EOS\space}%
	\providecommand \EOS [0]{\spacefactor3000\relax}%
	\providecommand \BibitemShut  [1]{\csname bibitem#1\endcsname}%
	\let\auto@bib@innerbib\@empty
	%</preamble>
	\bibitem [{\citenamefont {Qing-Yan}\ \emph {et~al.}(2012)\citenamefont
		{Qing-Yan}, \citenamefont {Zhi}, \citenamefont {Wen-Hao}, \citenamefont
		{Zuo-Cheng}, \citenamefont {Jin-Song}, \citenamefont {Wei}, \citenamefont
		{Hao}, \citenamefont {Yun-Bo}, \citenamefont {Peng}, \citenamefont {Kai},
		\citenamefont {Jing}, \citenamefont {Can-Li}, \citenamefont {Ke},
		\citenamefont {Jin-Feng}, \citenamefont {Shuai-Hua}, \citenamefont {Ya-Yu},
		\citenamefont {Li-Li}, \citenamefont {Xi}, \citenamefont {Xu-Cun},\ and\
		\citenamefont {Qi-Kun}}]{Qing-Yan2012}%
	\BibitemOpen
	\bibfield  {author} {\bibinfo {author} {\bibfnamefont {W.}~\bibnamefont
			{Qing-Yan}}, \bibinfo {author} {\bibfnamefont {L.}~\bibnamefont {Zhi}},
		\bibinfo {author} {\bibfnamefont {Z.}~\bibnamefont {Wen-Hao}}, \bibinfo
		{author} {\bibfnamefont {Z.}~\bibnamefont {Zuo-Cheng}}, \bibinfo {author}
		{\bibfnamefont {Z.}~\bibnamefont {Jin-Song}}, \bibinfo {author}
		{\bibfnamefont {L.}~\bibnamefont {Wei}}, \bibinfo {author} {\bibfnamefont
			{D.}~\bibnamefont {Hao}}, \bibinfo {author} {\bibfnamefont {O.}~\bibnamefont
			{Yun-Bo}}, \bibinfo {author} {\bibfnamefont {D.}~\bibnamefont {Peng}},
		\bibinfo {author} {\bibfnamefont {C.}~\bibnamefont {Kai}}, \bibinfo {author}
		{\bibfnamefont {W.}~\bibnamefont {Jing}}, \bibinfo {author} {\bibfnamefont
			{S.}~\bibnamefont {Can-Li}}, \bibinfo {author} {\bibfnamefont
			{H.}~\bibnamefont {Ke}}, \bibinfo {author} {\bibfnamefont {J.}~\bibnamefont
			{Jin-Feng}}, \bibinfo {author} {\bibfnamefont {J.}~\bibnamefont {Shuai-Hua}},
		\bibinfo {author} {\bibfnamefont {W.}~\bibnamefont {Ya-Yu}}, \bibinfo
		{author} {\bibfnamefont {W.}~\bibnamefont {Li-Li}}, \bibinfo {author}
		{\bibfnamefont {C.}~\bibnamefont {Xi}}, \bibinfo {author} {\bibfnamefont
			{M.}~\bibnamefont {Xu-Cun}}, \ and\ \bibinfo {author} {\bibfnamefont
			{X.}~\bibnamefont {Qi-Kun}},\ }\href
	{http://stacks.iop.org/0256-307X/29/i=3/a=037402} {\bibfield  {journal}
		{\bibinfo  {journal} {Chin. Phys. Lett.}\ }\textbf {\bibinfo {volume} {29}},\
		\bibinfo {pages} {037402} (\bibinfo {year} {2012})}\BibitemShut {NoStop}%
	\bibitem [{\citenamefont {Kreisel}\ \emph {et~al.}(2020)\citenamefont
		{Kreisel}, \citenamefont {Hirschfeld},\ and\ \citenamefont
		{Andersen}}]{Kreisel2020}%
	\BibitemOpen
	\bibfield  {author} {\bibinfo {author} {\bibfnamefont {A.}~\bibnamefont
			{Kreisel}}, \bibinfo {author} {\bibfnamefont {P.~J.}\ \bibnamefont
			{Hirschfeld}}, \ and\ \bibinfo {author} {\bibfnamefont {B.~M.}\ \bibnamefont
			{Andersen}},\ }\href@noop {} {\bibfield  {journal} {\bibinfo  {journal}
			{Symmetry}\ }\textbf {\bibinfo {volume} {12}},\ \bibinfo {pages} {1402}
		(\bibinfo {year} {2020})}\BibitemShut {NoStop}%
	\bibitem [{\citenamefont {Liu}\ \emph {et~al.}(2012)\citenamefont {Liu},
		\citenamefont {Zhang}, \citenamefont {Mou}, \citenamefont {He}, \citenamefont
		{Ou}, \citenamefont {Wang}, \citenamefont {Li}, \citenamefont {Wang},
		\citenamefont {Zhao}, \citenamefont {He}, \citenamefont {Peng}, \citenamefont
		{Liu}, \citenamefont {Chen}, \citenamefont {Yu}, \citenamefont {Liu},
		\citenamefont {Dong}, \citenamefont {Zhang}, \citenamefont {Chen},
		\citenamefont {Xu}, \citenamefont {Hu}, \citenamefont {Chen}, \citenamefont
		{Ma}, \citenamefont {Xue},\ and\ \citenamefont {Zhou}}]{Liu2012}%
	\BibitemOpen
	\bibfield  {author} {\bibinfo {author} {\bibfnamefont {D.}~\bibnamefont
			{Liu}}, \bibinfo {author} {\bibfnamefont {W.}~\bibnamefont {Zhang}}, \bibinfo
		{author} {\bibfnamefont {D.}~\bibnamefont {Mou}}, \bibinfo {author}
		{\bibfnamefont {J.}~\bibnamefont {He}}, \bibinfo {author} {\bibfnamefont
			{Y.-B.}\ \bibnamefont {Ou}}, \bibinfo {author} {\bibfnamefont {Q.-Y.}\
			\bibnamefont {Wang}}, \bibinfo {author} {\bibfnamefont {Z.}~\bibnamefont
			{Li}}, \bibinfo {author} {\bibfnamefont {L.}~\bibnamefont {Wang}}, \bibinfo
		{author} {\bibfnamefont {L.}~\bibnamefont {Zhao}}, \bibinfo {author}
		{\bibfnamefont {S.}~\bibnamefont {He}}, \bibinfo {author} {\bibfnamefont
			{Y.}~\bibnamefont {Peng}}, \bibinfo {author} {\bibfnamefont {X.}~\bibnamefont
			{Liu}}, \bibinfo {author} {\bibfnamefont {C.}~\bibnamefont {Chen}}, \bibinfo
		{author} {\bibfnamefont {L.}~\bibnamefont {Yu}}, \bibinfo {author}
		{\bibfnamefont {G.}~\bibnamefont {Liu}}, \bibinfo {author} {\bibfnamefont
			{X.}~\bibnamefont {Dong}}, \bibinfo {author} {\bibfnamefont {J.}~\bibnamefont
			{Zhang}}, \bibinfo {author} {\bibfnamefont {C.}~\bibnamefont {Chen}},
		\bibinfo {author} {\bibfnamefont {Z.}~\bibnamefont {Xu}}, \bibinfo {author}
		{\bibfnamefont {J.}~\bibnamefont {Hu}}, \bibinfo {author} {\bibfnamefont
			{X.}~\bibnamefont {Chen}}, \bibinfo {author} {\bibfnamefont {X.}~\bibnamefont
			{Ma}}, \bibinfo {author} {\bibfnamefont {Q.}~\bibnamefont {Xue}}, \ and\
		\bibinfo {author} {\bibfnamefont {X.}~\bibnamefont {Zhou}},\ }\href
	{http://dx.doi.org/10.1038/ncomms1946} {\bibfield  {journal} {\bibinfo
			{journal} {Nat. Commun.}\ }\textbf {\bibinfo {volume} {3}},\ \bibinfo {pages}
		{931} (\bibinfo {year} {2012})}\BibitemShut {NoStop}%
	\bibitem [{\citenamefont {He}\ \emph {et~al.}(2013)\citenamefont {He},
		\citenamefont {He}, \citenamefont {Zhang}, \citenamefont {Zhao},
		\citenamefont {Liu}, \citenamefont {Liu}, \citenamefont {Mou}, \citenamefont
		{Ou}, \citenamefont {Wang}, \citenamefont {Li}, \citenamefont {Wang},
		\citenamefont {Peng}, \citenamefont {Liu}, \citenamefont {Chen},
		\citenamefont {Yu}, \citenamefont {Liu}, \citenamefont {Dong}, \citenamefont
		{Zhang}, \citenamefont {Chen}, \citenamefont {Xu}, \citenamefont {Chen},
		\citenamefont {Ma}, \citenamefont {Xue},\ and\ \citenamefont
		{Zhou}}]{He2013}%
	\BibitemOpen
	\bibfield  {author} {\bibinfo {author} {\bibfnamefont {S.}~\bibnamefont
			{He}}, \bibinfo {author} {\bibfnamefont {J.}~\bibnamefont {He}}, \bibinfo
		{author} {\bibfnamefont {W.}~\bibnamefont {Zhang}}, \bibinfo {author}
		{\bibfnamefont {L.}~\bibnamefont {Zhao}}, \bibinfo {author} {\bibfnamefont
			{D.}~\bibnamefont {Liu}}, \bibinfo {author} {\bibfnamefont {X.}~\bibnamefont
			{Liu}}, \bibinfo {author} {\bibfnamefont {D.}~\bibnamefont {Mou}}, \bibinfo
		{author} {\bibfnamefont {Y.-B.}\ \bibnamefont {Ou}}, \bibinfo {author}
		{\bibfnamefont {Q.-Y.}\ \bibnamefont {Wang}}, \bibinfo {author}
		{\bibfnamefont {Z.}~\bibnamefont {Li}}, \bibinfo {author} {\bibfnamefont
			{L.}~\bibnamefont {Wang}}, \bibinfo {author} {\bibfnamefont {Y.}~\bibnamefont
			{Peng}}, \bibinfo {author} {\bibfnamefont {Y.}~\bibnamefont {Liu}}, \bibinfo
		{author} {\bibfnamefont {C.}~\bibnamefont {Chen}}, \bibinfo {author}
		{\bibfnamefont {L.}~\bibnamefont {Yu}}, \bibinfo {author} {\bibfnamefont
			{G.}~\bibnamefont {Liu}}, \bibinfo {author} {\bibfnamefont {X.}~\bibnamefont
			{Dong}}, \bibinfo {author} {\bibfnamefont {J.}~\bibnamefont {Zhang}},
		\bibinfo {author} {\bibfnamefont {C.}~\bibnamefont {Chen}}, \bibinfo {author}
		{\bibfnamefont {Z.}~\bibnamefont {Xu}}, \bibinfo {author} {\bibfnamefont
			{X.}~\bibnamefont {Chen}}, \bibinfo {author} {\bibfnamefont {X.}~\bibnamefont
			{Ma}}, \bibinfo {author} {\bibfnamefont {Q.}~\bibnamefont {Xue}}, \ and\
		\bibinfo {author} {\bibfnamefont {X.~J.}\ \bibnamefont {Zhou}},\ }\href
	{http://dx.doi.org/10.1038/nmat3648} {\bibfield  {journal} {\bibinfo
			{journal} {Nat. Mater.}\ }\textbf {\bibinfo {volume} {12}},\ \bibinfo {pages}
		{605} (\bibinfo {year} {2013})}\BibitemShut {NoStop}%
	\bibitem [{\citenamefont {Tan}\ \emph {et~al.}(2013)\citenamefont {Tan},
		\citenamefont {Zhang}, \citenamefont {Xia}, \citenamefont {Ye}, \citenamefont
		{Chen}, \citenamefont {Xie}, \citenamefont {Peng}, \citenamefont {Xu},
		\citenamefont {Fan}, \citenamefont {Xu}, \citenamefont {Jiang}, \citenamefont
		{Zhang}, \citenamefont {Lai}, \citenamefont {Xiang}, \citenamefont {Hu},
		\citenamefont {Xie},\ and\ \citenamefont {Feng}}]{Tan2013}%
	\BibitemOpen
	\bibfield  {author} {\bibinfo {author} {\bibfnamefont {S.}~\bibnamefont
			{Tan}}, \bibinfo {author} {\bibfnamefont {Y.}~\bibnamefont {Zhang}}, \bibinfo
		{author} {\bibfnamefont {M.}~\bibnamefont {Xia}}, \bibinfo {author}
		{\bibfnamefont {Z.}~\bibnamefont {Ye}}, \bibinfo {author} {\bibfnamefont
			{F.}~\bibnamefont {Chen}}, \bibinfo {author} {\bibfnamefont {X.}~\bibnamefont
			{Xie}}, \bibinfo {author} {\bibfnamefont {R.}~\bibnamefont {Peng}}, \bibinfo
		{author} {\bibfnamefont {D.}~\bibnamefont {Xu}}, \bibinfo {author}
		{\bibfnamefont {Q.}~\bibnamefont {Fan}}, \bibinfo {author} {\bibfnamefont
			{H.}~\bibnamefont {Xu}}, \bibinfo {author} {\bibfnamefont {J.}~\bibnamefont
			{Jiang}}, \bibinfo {author} {\bibfnamefont {T.}~\bibnamefont {Zhang}},
		\bibinfo {author} {\bibfnamefont {X.}~\bibnamefont {Lai}}, \bibinfo {author}
		{\bibfnamefont {T.}~\bibnamefont {Xiang}}, \bibinfo {author} {\bibfnamefont
			{J.}~\bibnamefont {Hu}}, \bibinfo {author} {\bibfnamefont {B.}~\bibnamefont
			{Xie}}, \ and\ \bibinfo {author} {\bibfnamefont {D.}~\bibnamefont {Feng}},\
	}\href {http://dx.doi.org/10.1038/nmat3654} {\bibfield  {journal} {\bibinfo
			{journal} {Nat. Mater.}\ }\textbf {\bibinfo {volume} {12}},\ \bibinfo {pages}
		{634} (\bibinfo {year} {2013})}\BibitemShut {NoStop}%
	\bibitem [{\citenamefont {Peng}\ \emph {et~al.}(2014)\citenamefont {Peng},
		\citenamefont {Shen}, \citenamefont {Xie}, \citenamefont {Xu}, \citenamefont
		{Tan}, \citenamefont {Xia}, \citenamefont {Zhang}, \citenamefont {Cao},
		\citenamefont {Gong}, \citenamefont {Hu}, \citenamefont {Xie},\ and\
		\citenamefont {Feng}}]{Peng2014}%
	\BibitemOpen
	\bibfield  {author} {\bibinfo {author} {\bibfnamefont {R.}~\bibnamefont
			{Peng}}, \bibinfo {author} {\bibfnamefont {X.~P.}\ \bibnamefont {Shen}},
		\bibinfo {author} {\bibfnamefont {X.}~\bibnamefont {Xie}}, \bibinfo {author}
		{\bibfnamefont {H.~C.}\ \bibnamefont {Xu}}, \bibinfo {author} {\bibfnamefont
			{S.~Y.}\ \bibnamefont {Tan}}, \bibinfo {author} {\bibfnamefont
			{M.}~\bibnamefont {Xia}}, \bibinfo {author} {\bibfnamefont {T.}~\bibnamefont
			{Zhang}}, \bibinfo {author} {\bibfnamefont {H.~Y.}\ \bibnamefont {Cao}},
		\bibinfo {author} {\bibfnamefont {X.~G.}\ \bibnamefont {Gong}}, \bibinfo
		{author} {\bibfnamefont {J.~P.}\ \bibnamefont {Hu}}, \bibinfo {author}
		{\bibfnamefont {B.~P.}\ \bibnamefont {Xie}}, \ and\ \bibinfo {author}
		{\bibfnamefont {D.~L.}\ \bibnamefont {Feng}},\ }\href {\doibase
		10.1103/PhysRevLett.112.107001} {\bibfield  {journal} {\bibinfo  {journal}
			{Phys. Rev. Lett.}\ }\textbf {\bibinfo {volume} {112}},\ \bibinfo {pages}
		{107001} (\bibinfo {year} {2014})}\BibitemShut {NoStop}%
	\bibitem [{\citenamefont {Lee}\ \emph {et~al.}(2014)\citenamefont {Lee},
		\citenamefont {Schmitt}, \citenamefont {Moore}, \citenamefont {Johnston},
		\citenamefont {Cui}, \citenamefont {Li}, \citenamefont {Yi}, \citenamefont
		{Liu}, \citenamefont {Hashimoto}, \citenamefont {Zhang}, \citenamefont {Lu},
		\citenamefont {Devereaux}, \citenamefont {Lee},\ and\ \citenamefont
		{Shen}}]{Lee2014}%
	\BibitemOpen
	\bibfield  {author} {\bibinfo {author} {\bibfnamefont {J.~J.}\ \bibnamefont
			{Lee}}, \bibinfo {author} {\bibfnamefont {F.~T.}\ \bibnamefont {Schmitt}},
		\bibinfo {author} {\bibfnamefont {R.~G.}\ \bibnamefont {Moore}}, \bibinfo
		{author} {\bibfnamefont {S.}~\bibnamefont {Johnston}}, \bibinfo {author}
		{\bibfnamefont {Y.-T.}\ \bibnamefont {Cui}}, \bibinfo {author} {\bibfnamefont
			{W.}~\bibnamefont {Li}}, \bibinfo {author} {\bibfnamefont {M.}~\bibnamefont
			{Yi}}, \bibinfo {author} {\bibfnamefont {Z.~K.}\ \bibnamefont {Liu}},
		\bibinfo {author} {\bibfnamefont {M.}~\bibnamefont {Hashimoto}}, \bibinfo
		{author} {\bibfnamefont {Y.}~\bibnamefont {Zhang}}, \bibinfo {author}
		{\bibfnamefont {D.~H.}\ \bibnamefont {Lu}}, \bibinfo {author} {\bibfnamefont
			{T.~P.}\ \bibnamefont {Devereaux}}, \bibinfo {author} {\bibfnamefont {D.-H.}\
			\bibnamefont {Lee}}, \ and\ \bibinfo {author} {\bibfnamefont {Z.-X.}\
			\bibnamefont {Shen}},\ }\href {http://dx.doi.org/10.1038/nature13894}
	{\bibfield  {journal} {\bibinfo  {journal} {Nature}\ }\textbf {\bibinfo
			{volume} {515}},\ \bibinfo {pages} {245} (\bibinfo {year}
		{2014})}\BibitemShut {NoStop}%
	\bibitem [{\citenamefont {Ge}\ \emph {et~al.}(2015)\citenamefont {Ge},
		\citenamefont {Liu}, \citenamefont {Liu}, \citenamefont {Gao}, \citenamefont
		{Qian}, \citenamefont {Xue}, \citenamefont {Liu},\ and\ \citenamefont
		{Jia}}]{Ge2015}%
	\BibitemOpen
	\bibfield  {author} {\bibinfo {author} {\bibfnamefont {J.-F.}\ \bibnamefont
			{Ge}}, \bibinfo {author} {\bibfnamefont {Z.-L.}\ \bibnamefont {Liu}},
		\bibinfo {author} {\bibfnamefont {C.}~\bibnamefont {Liu}}, \bibinfo {author}
		{\bibfnamefont {C.-L.}\ \bibnamefont {Gao}}, \bibinfo {author} {\bibfnamefont
			{D.}~\bibnamefont {Qian}}, \bibinfo {author} {\bibfnamefont {Q.-K.}\
			\bibnamefont {Xue}}, \bibinfo {author} {\bibfnamefont {Y.}~\bibnamefont
			{Liu}}, \ and\ \bibinfo {author} {\bibfnamefont {J.-F.}\ \bibnamefont
			{Jia}},\ }\href {http://dx.doi.org/10.1038/nmat4153} {\bibfield  {journal}
		{\bibinfo  {journal} {Nat. Mater.}\ }\textbf {\bibinfo {volume} {14}},\
		\bibinfo {pages} {285} (\bibinfo {year} {2015})}\BibitemShut {NoStop}%
	\bibitem [{\citenamefont {Hsu}\ \emph {et~al.}(2008)\citenamefont {Hsu},
		\citenamefont {Luo}, \citenamefont {Yeh}, \citenamefont {Chen}, \citenamefont
		{Huang}, \citenamefont {Wu}, \citenamefont {Lee}, \citenamefont {Huang},
		\citenamefont {Chu}, \citenamefont {Yan},\ and\ \citenamefont
		{Wu}}]{Hsu2008}%
	\BibitemOpen
	\bibfield  {author} {\bibinfo {author} {\bibfnamefont {F.-C.}\ \bibnamefont
			{Hsu}}, \bibinfo {author} {\bibfnamefont {J.-Y.}\ \bibnamefont {Luo}},
		\bibinfo {author} {\bibfnamefont {K.-W.}\ \bibnamefont {Yeh}}, \bibinfo
		{author} {\bibfnamefont {T.-K.}\ \bibnamefont {Chen}}, \bibinfo {author}
		{\bibfnamefont {T.-W.}\ \bibnamefont {Huang}}, \bibinfo {author}
		{\bibfnamefont {P.~M.}\ \bibnamefont {Wu}}, \bibinfo {author} {\bibfnamefont
			{Y.-C.}\ \bibnamefont {Lee}}, \bibinfo {author} {\bibfnamefont {Y.-L.}\
			\bibnamefont {Huang}}, \bibinfo {author} {\bibfnamefont {Y.-Y.}\ \bibnamefont
			{Chu}}, \bibinfo {author} {\bibfnamefont {D.-C.}\ \bibnamefont {Yan}}, \ and\
		\bibinfo {author} {\bibfnamefont {M.-K.}\ \bibnamefont {Wu}},\ }\href
	{\doibase 10.1073/pnas.0807325105} {\bibfield  {journal} {\bibinfo  {journal}
			{Proc. Natl. Acad. Sci. USA}\ }\textbf {\bibinfo {volume} {105}},\ \bibinfo
		{pages} {14262} (\bibinfo {year} {2008})}\BibitemShut {NoStop}%
	\bibitem [{\citenamefont {Imai}\ \emph {et~al.}(2009)\citenamefont {Imai},
		\citenamefont {Ahilan}, \citenamefont {Ning}, \citenamefont {McQueen},\ and\
		\citenamefont {Cava}}]{Imai2009}%
	\BibitemOpen
	\bibfield  {author} {\bibinfo {author} {\bibfnamefont {T.}~\bibnamefont
			{Imai}}, \bibinfo {author} {\bibfnamefont {K.}~\bibnamefont {Ahilan}},
		\bibinfo {author} {\bibfnamefont {F.~L.}\ \bibnamefont {Ning}}, \bibinfo
		{author} {\bibfnamefont {T.~M.}\ \bibnamefont {McQueen}}, \ and\ \bibinfo
		{author} {\bibfnamefont {R.~J.}\ \bibnamefont {Cava}},\ }\href {\doibase
		10.1103/PhysRevLett.102.177005} {\bibfield  {journal} {\bibinfo  {journal}
			{Phys. Rev. Lett.}\ }\textbf {\bibinfo {volume} {102}},\ \bibinfo {pages}
		{177005} (\bibinfo {year} {2009})}\BibitemShut {NoStop}%
	\bibitem [{\citenamefont {Rahn}\ \emph {et~al.}(2015)\citenamefont {Rahn},
		\citenamefont {Ewings}, \citenamefont {Sedlmaier}, \citenamefont {Clarke},\
		and\ \citenamefont {Boothroyd}}]{Rahn2015}%
	\BibitemOpen
	\bibfield  {author} {\bibinfo {author} {\bibfnamefont {M.~C.}\ \bibnamefont
			{Rahn}}, \bibinfo {author} {\bibfnamefont {R.~A.}\ \bibnamefont {Ewings}},
		\bibinfo {author} {\bibfnamefont {S.~J.}\ \bibnamefont {Sedlmaier}}, \bibinfo
		{author} {\bibfnamefont {S.~J.}\ \bibnamefont {Clarke}}, \ and\ \bibinfo
		{author} {\bibfnamefont {A.~T.}\ \bibnamefont {Boothroyd}},\ }\href@noop {}
	{\bibfield  {journal} {\bibinfo  {journal} {Phys. Rev. B}\ }\textbf {\bibinfo
			{volume} {91}},\ \bibinfo {pages} {180501} (\bibinfo {year}
		{2015})}\BibitemShut {NoStop}%
	\bibitem [{\citenamefont {Wang}\ \emph
		{et~al.}(2016{\natexlab{a}})\citenamefont {Wang}, \citenamefont {Shen},
		\citenamefont {Pan}, \citenamefont {Hao}, \citenamefont {Ma}, \citenamefont
		{Zhou}, \citenamefont {Steffens}, \citenamefont {Schmalzl}, \citenamefont
		{Forrest}, \citenamefont {Abdel-Hafiez}, \citenamefont {Chareev},
		\citenamefont {Vasiliev}, \citenamefont {Bourges}, \citenamefont {Sidis},
		\citenamefont {Cao},\ and\ \citenamefont {Zhao}}]{Wang2016NatMat}%
	\BibitemOpen
	\bibfield  {author} {\bibinfo {author} {\bibfnamefont {Q.}~\bibnamefont
			{Wang}}, \bibinfo {author} {\bibfnamefont {Y.}~\bibnamefont {Shen}}, \bibinfo
		{author} {\bibfnamefont {B.}~\bibnamefont {Pan}}, \bibinfo {author}
		{\bibfnamefont {Y.}~\bibnamefont {Hao}}, \bibinfo {author} {\bibfnamefont
			{M.}~\bibnamefont {Ma}}, \bibinfo {author} {\bibfnamefont {F.}~\bibnamefont
			{Zhou}}, \bibinfo {author} {\bibfnamefont {P.}~\bibnamefont {Steffens}},
		\bibinfo {author} {\bibfnamefont {K.}~\bibnamefont {Schmalzl}}, \bibinfo
		{author} {\bibfnamefont {T.~R.}\ \bibnamefont {Forrest}}, \bibinfo {author}
		{\bibfnamefont {M.}~\bibnamefont {Abdel-Hafiez}}, \bibinfo {author}
		{\bibfnamefont {D.~A.}\ \bibnamefont {Chareev}}, \bibinfo {author}
		{\bibfnamefont {A.~N.}\ \bibnamefont {Vasiliev}}, \bibinfo {author}
		{\bibfnamefont {P.}~\bibnamefont {Bourges}}, \bibinfo {author} {\bibfnamefont
			{Y.}~\bibnamefont {Sidis}}, \bibinfo {author} {\bibfnamefont
			{H.}~\bibnamefont {Cao}}, \ and\ \bibinfo {author} {\bibfnamefont
			{J.}~\bibnamefont {Zhao}},\ }\href@noop {} {\bibfield  {journal} {\bibinfo
			{journal} {Nat. Mater.}\ }\textbf {\bibinfo {volume} {15}},\ \bibinfo {pages}
		{159 } (\bibinfo {year} {2016}{\natexlab{a}})}\BibitemShut {NoStop}%
	\bibitem [{\citenamefont {Wang}\ \emph
		{et~al.}(2016{\natexlab{b}})\citenamefont {Wang}, \citenamefont {Shen},
		\citenamefont {Pan}, \citenamefont {Zhang}, \citenamefont {Ikeuchi},
		\citenamefont {Iida}, \citenamefont {Christianson}, \citenamefont {Walker},
		\citenamefont {Adroja}, \citenamefont {Abdel-Hafiez}, \citenamefont {Chen},
		\citenamefont {Chareev}, \citenamefont {Vasiliev},\ and\ \citenamefont
		{Zhao}}]{Wang2016NatCom}%
	\BibitemOpen
	\bibfield  {author} {\bibinfo {author} {\bibfnamefont {Q.}~\bibnamefont
			{Wang}}, \bibinfo {author} {\bibfnamefont {Y.}~\bibnamefont {Shen}}, \bibinfo
		{author} {\bibfnamefont {B.}~\bibnamefont {Pan}}, \bibinfo {author}
		{\bibfnamefont {X.}~\bibnamefont {Zhang}}, \bibinfo {author} {\bibfnamefont
			{K.}~\bibnamefont {Ikeuchi}}, \bibinfo {author} {\bibfnamefont
			{K.}~\bibnamefont {Iida}}, \bibinfo {author} {\bibfnamefont {A.~D.}\
			\bibnamefont {Christianson}}, \bibinfo {author} {\bibfnamefont {H.~C.}\
			\bibnamefont {Walker}}, \bibinfo {author} {\bibfnamefont {D.~T.}\
			\bibnamefont {Adroja}}, \bibinfo {author} {\bibfnamefont {M.}~\bibnamefont
			{Abdel-Hafiez}}, \bibinfo {author} {\bibfnamefont {X.}~\bibnamefont {Chen}},
		\bibinfo {author} {\bibfnamefont {D.~A.}\ \bibnamefont {Chareev}}, \bibinfo
		{author} {\bibfnamefont {A.~N.}\ \bibnamefont {Vasiliev}}, \ and\ \bibinfo
		{author} {\bibfnamefont {J.}~\bibnamefont {Zhao}},\ }\href@noop {} {\bibfield
		{journal} {\bibinfo  {journal} {Nat. Commun.}\ }\textbf {\bibinfo {volume}
			{7}},\ \bibinfo {pages} {12182} (\bibinfo {year}
		{2016}{\natexlab{b}})}\BibitemShut {NoStop}%
	\bibitem [{\citenamefont {Baek}\ \emph {et~al.}(2020)\citenamefont {Baek},
		\citenamefont {Mok}, \citenamefont {Kim}, \citenamefont {Aswartham},
		\citenamefont {Morozov}, \citenamefont {Chareev}, \citenamefont {Urata},
		\citenamefont {Tanigaki}, \citenamefont {Tanabe}, \citenamefont
		{B{\"u}chner},\ and\ \citenamefont {Efremov}}]{Baek2020}%
	\BibitemOpen
	\bibfield  {author} {\bibinfo {author} {\bibfnamefont {S.-H.}\ \bibnamefont
			{Baek}}, \bibinfo {author} {\bibfnamefont {J.}~\bibnamefont {Mok}}, \bibinfo
		{author} {\bibfnamefont {J.~S.}\ \bibnamefont {Kim}}, \bibinfo {author}
		{\bibfnamefont {S.}~\bibnamefont {Aswartham}}, \bibinfo {author}
		{\bibfnamefont {I.}~\bibnamefont {Morozov}}, \bibinfo {author} {\bibfnamefont
			{D.}~\bibnamefont {Chareev}}, \bibinfo {author} {\bibfnamefont
			{T.}~\bibnamefont {Urata}}, \bibinfo {author} {\bibfnamefont
			{K.}~\bibnamefont {Tanigaki}}, \bibinfo {author} {\bibfnamefont
			{Y.}~\bibnamefont {Tanabe}}, \bibinfo {author} {\bibfnamefont
			{B.}~\bibnamefont {B{\"u}chner}}, \ and\ \bibinfo {author} {\bibfnamefont
			{D.~V.}\ \bibnamefont {Efremov}},\ }\href@noop {} {\bibfield  {journal}
		{\bibinfo  {journal} {njp Quant. Mater.}\ }\textbf {\bibinfo {volume} {5}},\
		\bibinfo {pages} {8} (\bibinfo {year} {2020})}\BibitemShut {NoStop}%
	\bibitem [{\citenamefont {Linscheid}\ \emph {et~al.}(2016)\citenamefont
		{Linscheid}, \citenamefont {Maiti}, \citenamefont {Wang}, \citenamefont
		{Johnston},\ and\ \citenamefont {Hirschfeld}}]{Linscheid2016a}%
	\BibitemOpen
	\bibfield  {author} {\bibinfo {author} {\bibfnamefont {A.}~\bibnamefont
			{Linscheid}}, \bibinfo {author} {\bibfnamefont {S.}~\bibnamefont {Maiti}},
		\bibinfo {author} {\bibfnamefont {Y.}~\bibnamefont {Wang}}, \bibinfo {author}
		{\bibfnamefont {S.}~\bibnamefont {Johnston}}, \ and\ \bibinfo {author}
		{\bibfnamefont {P.~J.}\ \bibnamefont {Hirschfeld}},\ }\href {\doibase
		10.1103/PhysRevLett.117.077003} {\bibfield  {journal} {\bibinfo  {journal}
			{Phys. Rev. Lett.}\ }\textbf {\bibinfo {volume} {117}},\ \bibinfo {pages}
		{077003} (\bibinfo {year} {2016})}\BibitemShut {NoStop}%
	\bibitem [{\citenamefont {Kreisel}\ \emph {et~al.}(2017)\citenamefont
		{Kreisel}, \citenamefont {Andersen}, \citenamefont {Sprau}, \citenamefont
		{Kostin}, \citenamefont {Davis},\ and\ \citenamefont
		{Hirschfeld}}]{Kreisel2017}%
	\BibitemOpen
	\bibfield  {author} {\bibinfo {author} {\bibfnamefont {A.}~\bibnamefont
			{Kreisel}}, \bibinfo {author} {\bibfnamefont {B.~M.}\ \bibnamefont
			{Andersen}}, \bibinfo {author} {\bibfnamefont {P.~O.}\ \bibnamefont {Sprau}},
		\bibinfo {author} {\bibfnamefont {A.}~\bibnamefont {Kostin}}, \bibinfo
		{author} {\bibfnamefont {J.~C.~S.}\ \bibnamefont {Davis}}, \ and\ \bibinfo
		{author} {\bibfnamefont {P.~J.}\ \bibnamefont {Hirschfeld}},\ }\href
	{\doibase 10.1103/PhysRevB.95.174504} {\bibfield  {journal} {\bibinfo
			{journal} {Phys. Rev. B}\ }\textbf {\bibinfo {volume} {95}},\ \bibinfo
		{pages} {174504} (\bibinfo {year} {2017})}\BibitemShut {NoStop}%
	\bibitem [{\citenamefont {Jandke}\ \emph {et~al.}(2019)\citenamefont {Jandke},
		\citenamefont {Yang}, \citenamefont {Hlobil}, \citenamefont {Engelhardt},
		\citenamefont {Rau}, \citenamefont {Zakeri}, \citenamefont {Gao},
		\citenamefont {Schmalian},\ and\ \citenamefont {Wulfhekel}}]{Jandke2019}%
	\BibitemOpen
	\bibfield  {author} {\bibinfo {author} {\bibfnamefont {J.}~\bibnamefont
			{Jandke}}, \bibinfo {author} {\bibfnamefont {F.}~\bibnamefont {Yang}},
		\bibinfo {author} {\bibfnamefont {P.}~\bibnamefont {Hlobil}}, \bibinfo
		{author} {\bibfnamefont {T.}~\bibnamefont {Engelhardt}}, \bibinfo {author}
		{\bibfnamefont {D.}~\bibnamefont {Rau}}, \bibinfo {author} {\bibfnamefont
			{K.}~\bibnamefont {Zakeri}}, \bibinfo {author} {\bibfnamefont
			{C.}~\bibnamefont {Gao}}, \bibinfo {author} {\bibfnamefont {J.}~\bibnamefont
			{Schmalian}}, \ and\ \bibinfo {author} {\bibfnamefont {W.}~\bibnamefont
			{Wulfhekel}},\ }\href {\doibase 10.1103/PhysRevB.100.020503} {\bibfield
		{journal} {\bibinfo  {journal} {Phys. Rev. B}\ }\textbf {\bibinfo {volume}
			{100}},\ \bibinfo {pages} {020503} (\bibinfo {year} {2019})}\BibitemShut
	{NoStop}%
	\bibitem [{\citenamefont {Liu}\ \emph {et~al.}(2019{\natexlab{a}})\citenamefont
		{Liu}, \citenamefont {Wang}, \citenamefont {Ye}, \citenamefont {Chen},
		\citenamefont {Liu}, \citenamefont {Wang}, \citenamefont {Wang},\ and\
		\citenamefont {Wang}}]{Liu2019a}%
	\BibitemOpen
	\bibfield  {author} {\bibinfo {author} {\bibfnamefont {C.}~\bibnamefont
			{Liu}}, \bibinfo {author} {\bibfnamefont {Z.}~\bibnamefont {Wang}}, \bibinfo
		{author} {\bibfnamefont {S.}~\bibnamefont {Ye}}, \bibinfo {author}
		{\bibfnamefont {C.}~\bibnamefont {Chen}}, \bibinfo {author} {\bibfnamefont
			{Y.}~\bibnamefont {Liu}}, \bibinfo {author} {\bibfnamefont {Q.}~\bibnamefont
			{Wang}}, \bibinfo {author} {\bibfnamefont {Q.-H.}\ \bibnamefont {Wang}}, \
		and\ \bibinfo {author} {\bibfnamefont {J.}~\bibnamefont {Wang}},\ }\href
	{\doibase 10.1021/acs.nanolett.9b00144} {\bibfield  {journal} {\bibinfo
			{journal} {Nano Letters}\ }\textbf {\bibinfo {volume} {19}},\ \bibinfo
		{pages} {3464} (\bibinfo {year} {2019}{\natexlab{a}})}\BibitemShut {NoStop}%
	\bibitem [{\citenamefont {Shishidou}\ \emph {et~al.}(2018)\citenamefont
		{Shishidou}, \citenamefont {Agterberg},\ and\ \citenamefont
		{Weinert}}]{Shishidou2018}%
	\BibitemOpen
	\bibfield  {author} {\bibinfo {author} {\bibfnamefont {T.}~\bibnamefont
			{Shishidou}}, \bibinfo {author} {\bibfnamefont {D.~F.}\ \bibnamefont
			{Agterberg}}, \ and\ \bibinfo {author} {\bibfnamefont {M.}~\bibnamefont
			{Weinert}},\ }\href {\doibase 10.1038/s42005-018-0006-7} {\bibfield
		{journal} {\bibinfo  {journal} {Commun. Phys.}\ }\textbf {\bibinfo {volume}
			{1}},\ \bibinfo {pages} {8} (\bibinfo {year} {2018})}\BibitemShut {NoStop}%
	\bibitem [{\citenamefont {Schrodi}\ \emph
		{et~al.}(2020{\natexlab{a}})\citenamefont {Schrodi}, \citenamefont {Aperis},\
		and\ \citenamefont {Oppeneer}}]{Schrodi2020_3}%
	\BibitemOpen
	\bibfield  {author} {\bibinfo {author} {\bibfnamefont {F.}~\bibnamefont
			{Schrodi}}, \bibinfo {author} {\bibfnamefont {A.}~\bibnamefont {Aperis}}, \
		and\ \bibinfo {author} {\bibfnamefont {P.~M.}\ \bibnamefont {Oppeneer}},\
	}\href {\doibase 10.1103/PhysRevB.102.014502} {\bibfield  {journal} {\bibinfo
			{journal} {Phys. Rev. B}\ }\textbf {\bibinfo {volume} {102}},\ \bibinfo
		{pages} {014502} (\bibinfo {year} {2020}{\natexlab{a}})}\BibitemShut
	{NoStop}%
	\bibitem [{\citenamefont {Rebec}\ \emph {et~al.}(2017)\citenamefont {Rebec},
		\citenamefont {Jia}, \citenamefont {Zhang}, \citenamefont {Hashimoto},
		\citenamefont {Lu}, \citenamefont {Moore},\ and\ \citenamefont
		{Shen}}]{Rebec2017}%
	\BibitemOpen
	\bibfield  {author} {\bibinfo {author} {\bibfnamefont {S.~N.}\ \bibnamefont
			{Rebec}}, \bibinfo {author} {\bibfnamefont {T.}~\bibnamefont {Jia}}, \bibinfo
		{author} {\bibfnamefont {C.}~\bibnamefont {Zhang}}, \bibinfo {author}
		{\bibfnamefont {M.}~\bibnamefont {Hashimoto}}, \bibinfo {author}
		{\bibfnamefont {D.-H.}\ \bibnamefont {Lu}}, \bibinfo {author} {\bibfnamefont
			{R.~G.}\ \bibnamefont {Moore}}, \ and\ \bibinfo {author} {\bibfnamefont
			{Z.-X.}\ \bibnamefont {Shen}},\ }\href {\doibase
		10.1103/PhysRevLett.118.067002} {\bibfield  {journal} {\bibinfo  {journal}
			{Phys. Rev. Lett.}\ }\textbf {\bibinfo {volume} {118}},\ \bibinfo {pages}
		{067002} (\bibinfo {year} {2017})}\BibitemShut {NoStop}%
	\bibitem [{\citenamefont {Li}\ \emph {et~al.}(2014)\citenamefont {Li},
		\citenamefont {Xing}, \citenamefont {Huang},\ and\ \citenamefont
		{Xing}}]{Li2014}%
	\BibitemOpen
	\bibfield  {author} {\bibinfo {author} {\bibfnamefont {B.}~\bibnamefont
			{Li}}, \bibinfo {author} {\bibfnamefont {Z.~W.}\ \bibnamefont {Xing}},
		\bibinfo {author} {\bibfnamefont {G.~Q.}\ \bibnamefont {Huang}}, \ and\
		\bibinfo {author} {\bibfnamefont {D.~Y.}\ \bibnamefont {Xing}},\ }\href
	{\doibase http://dx.doi.org/10.1063/1.4876750} {\bibfield  {journal}
		{\bibinfo  {journal} {J. Appl. Phys.}\ }\textbf {\bibinfo {volume} {115}},\
		\bibinfo {eid} {193907} (\bibinfo {year} {2014})}\BibitemShut {NoStop}%
	\bibitem [{\citenamefont {Xie}\ \emph {et~al.}(2015)\citenamefont {Xie},
		\citenamefont {Cao}, \citenamefont {Zhou}, \citenamefont {Chen},
		\citenamefont {Xiang},\ and\ \citenamefont {Gong}}]{Xie2015}%
	\BibitemOpen
	\bibfield  {author} {\bibinfo {author} {\bibfnamefont {Y.}~\bibnamefont
			{Xie}}, \bibinfo {author} {\bibfnamefont {H.-Y.}\ \bibnamefont {Cao}},
		\bibinfo {author} {\bibfnamefont {Y.}~\bibnamefont {Zhou}}, \bibinfo {author}
		{\bibfnamefont {S.}~\bibnamefont {Chen}}, \bibinfo {author} {\bibfnamefont
			{H.}~\bibnamefont {Xiang}}, \ and\ \bibinfo {author} {\bibfnamefont {X.-G.}\
			\bibnamefont {Gong}},\ }\href {http://dx.doi.org/10.1038/srep10011}
	{\bibfield  {journal} {\bibinfo  {journal} {Sci. Rep.}\ }\textbf {\bibinfo
			{volume} {5}},\ \bibinfo {pages} {10011} (\bibinfo {year}
		{2015})}\BibitemShut {NoStop}%
	\bibitem [{\citenamefont {Wang}\ \emph
		{et~al.}(2016{\natexlab{c}})\citenamefont {Wang}, \citenamefont {Linscheid},
		\citenamefont {Berlijn},\ and\ \citenamefont {Johnston}}]{Wang2016b}%
	\BibitemOpen
	\bibfield  {author} {\bibinfo {author} {\bibfnamefont {Y.}~\bibnamefont
			{Wang}}, \bibinfo {author} {\bibfnamefont {A.}~\bibnamefont {Linscheid}},
		\bibinfo {author} {\bibfnamefont {T.}~\bibnamefont {Berlijn}}, \ and\
		\bibinfo {author} {\bibfnamefont {S.}~\bibnamefont {Johnston}},\ }\href
	{\doibase 10.1103/PhysRevB.93.134513} {\bibfield  {journal} {\bibinfo
			{journal} {Phys. Rev. B}\ }\textbf {\bibinfo {volume} {93}},\ \bibinfo
		{pages} {134513} (\bibinfo {year} {2016}{\natexlab{c}})}\BibitemShut
	{NoStop}%
	\bibitem [{\citenamefont {Zhou}\ and\ \citenamefont
		{Millis}(2016)}]{Zhou2016a}%
	\BibitemOpen
	\bibfield  {author} {\bibinfo {author} {\bibfnamefont {Y.}~\bibnamefont
			{Zhou}}\ and\ \bibinfo {author} {\bibfnamefont {A.~J.}\ \bibnamefont
			{Millis}},\ }\href {\doibase 10.1103/PhysRevB.93.224506} {\bibfield
		{journal} {\bibinfo  {journal} {Phys. Rev. B}\ }\textbf {\bibinfo {volume}
			{93}},\ \bibinfo {pages} {224506} (\bibinfo {year} {2016})}\BibitemShut
	{NoStop}%
	\bibitem [{\citenamefont {Rademaker}\ \emph {et~al.}(2016)\citenamefont
		{Rademaker}, \citenamefont {Wang}, \citenamefont {Berlijn},\ and\
		\citenamefont {Johnston}}]{Rademaker2016}%
	\BibitemOpen
	\bibfield  {author} {\bibinfo {author} {\bibfnamefont {L.}~\bibnamefont
			{Rademaker}}, \bibinfo {author} {\bibfnamefont {Y.}~\bibnamefont {Wang}},
		\bibinfo {author} {\bibfnamefont {T.}~\bibnamefont {Berlijn}}, \ and\
		\bibinfo {author} {\bibfnamefont {S.}~\bibnamefont {Johnston}},\ }\href
	{http://stacks.iop.org/1367-2630/18/i=2/a=022001} {\bibfield  {journal}
		{\bibinfo  {journal} {New J. Phys.}\ }\textbf {\bibinfo {volume} {18}},\
		\bibinfo {pages} {022001} (\bibinfo {year} {2016})}\BibitemShut {NoStop}%
	\bibitem [{\citenamefont {Wang}\ \emph
		{et~al.}(2016{\natexlab{d}})\citenamefont {Wang}, \citenamefont
		{Nakatsukasa}, \citenamefont {Rademaker}, \citenamefont {Berlijn},\ and\
		\citenamefont {Johnston}}]{Wang2016c}%
	\BibitemOpen
	\bibfield  {author} {\bibinfo {author} {\bibfnamefont {Y.}~\bibnamefont
			{Wang}}, \bibinfo {author} {\bibfnamefont {K.}~\bibnamefont {Nakatsukasa}},
		\bibinfo {author} {\bibfnamefont {L.}~\bibnamefont {Rademaker}}, \bibinfo
		{author} {\bibfnamefont {T.}~\bibnamefont {Berlijn}}, \ and\ \bibinfo
		{author} {\bibfnamefont {S.}~\bibnamefont {Johnston}},\ }\href@noop {}
	{\bibfield  {journal} {\bibinfo  {journal} {Supercond. Sci. Technol.}\
		}\textbf {\bibinfo {volume} {29}},\ \bibinfo {pages} {054009} (\bibinfo
		{year} {2016}{\natexlab{d}})}\BibitemShut {NoStop}%
	\bibitem [{\citenamefont {Aperis}\ and\ \citenamefont
		{Oppeneer}(2018)}]{Aperis2018}%
	\BibitemOpen
	\bibfield  {author} {\bibinfo {author} {\bibfnamefont {A.}~\bibnamefont
			{Aperis}}\ and\ \bibinfo {author} {\bibfnamefont {P.~M.}\ \bibnamefont
			{Oppeneer}},\ }\href {\doibase 10.1103/PhysRevB.97.060501} {\bibfield
		{journal} {\bibinfo  {journal} {Phys. Rev. B}\ }\textbf {\bibinfo {volume}
			{97}},\ \bibinfo {pages} {060501} (\bibinfo {year} {2018})}\BibitemShut
	{NoStop}%
	\bibitem [{\citenamefont {Schrodi}\ \emph {et~al.}(2018)\citenamefont
		{Schrodi}, \citenamefont {Aperis},\ and\ \citenamefont
		{Oppeneer}}]{Schrodi2018}%
	\BibitemOpen
	\bibfield  {author} {\bibinfo {author} {\bibfnamefont {F.}~\bibnamefont
			{Schrodi}}, \bibinfo {author} {\bibfnamefont {A.}~\bibnamefont {Aperis}}, \
		and\ \bibinfo {author} {\bibfnamefont {P.~M.}\ \bibnamefont {Oppeneer}},\
	}\href {\doibase 10.1103/PhysRevB.98.094509} {\bibfield  {journal} {\bibinfo
			{journal} {Phys. Rev. B}\ }\textbf {\bibinfo {volume} {98}},\ \bibinfo
		{pages} {094509} (\bibinfo {year} {2018})}\BibitemShut {NoStop}%
	\bibitem [{\citenamefont {Song}\ \emph {et~al.}(2019)\citenamefont {Song},
		\citenamefont {Yu}, \citenamefont {Lou}, \citenamefont {Xie}, \citenamefont
		{Xu}, \citenamefont {Wen}, \citenamefont {Yao}, \citenamefont {Zhang},
		\citenamefont {Zhu}, \citenamefont {Guo}, \citenamefont {Peng},\ and\
		\citenamefont {Feng}}]{Song2019}%
	\BibitemOpen
	\bibfield  {author} {\bibinfo {author} {\bibfnamefont {Q.}~\bibnamefont
			{Song}}, \bibinfo {author} {\bibfnamefont {T.~L.}\ \bibnamefont {Yu}},
		\bibinfo {author} {\bibfnamefont {X.}~\bibnamefont {Lou}}, \bibinfo {author}
		{\bibfnamefont {B.~P.}\ \bibnamefont {Xie}}, \bibinfo {author} {\bibfnamefont
			{H.~C.}\ \bibnamefont {Xu}}, \bibinfo {author} {\bibfnamefont {C.~H.~P.}\
			\bibnamefont {Wen}}, \bibinfo {author} {\bibfnamefont {Q.}~\bibnamefont
			{Yao}}, \bibinfo {author} {\bibfnamefont {S.~Y.}\ \bibnamefont {Zhang}},
		\bibinfo {author} {\bibfnamefont {X.~T.}\ \bibnamefont {Zhu}}, \bibinfo
		{author} {\bibfnamefont {J.~D.}\ \bibnamefont {Guo}}, \bibinfo {author}
		{\bibfnamefont {R.}~\bibnamefont {Peng}}, \ and\ \bibinfo {author}
		{\bibfnamefont {D.~L.}\ \bibnamefont {Feng}},\ }\href {\doibase
		10.1038/s41467-019-08560-z} {\bibfield  {journal} {\bibinfo  {journal} {Nat.
				Commun.}\ }\textbf {\bibinfo {volume} {10}},\ \bibinfo {pages} {758}
		(\bibinfo {year} {2019})}\BibitemShut {NoStop}%
	\bibitem [{\citenamefont {Fan}\ \emph {et~al.}(2015)\citenamefont {Fan},
		\citenamefont {Zhang}, \citenamefont {Liu}, \citenamefont {Yan},
		\citenamefont {Ren}, \citenamefont {Peng}, \citenamefont {Xu}, \citenamefont
		{Xie}, \citenamefont {Hu}, \citenamefont {Zhang},\ and\ \citenamefont
		{Feng}}]{Fan2015}%
	\BibitemOpen
	\bibfield  {author} {\bibinfo {author} {\bibfnamefont {Q.}~\bibnamefont
			{Fan}}, \bibinfo {author} {\bibfnamefont {W.~H.}\ \bibnamefont {Zhang}},
		\bibinfo {author} {\bibfnamefont {X.}~\bibnamefont {Liu}}, \bibinfo {author}
		{\bibfnamefont {Y.~J.}\ \bibnamefont {Yan}}, \bibinfo {author} {\bibfnamefont
			{M.~Q.}\ \bibnamefont {Ren}}, \bibinfo {author} {\bibfnamefont
			{R.}~\bibnamefont {Peng}}, \bibinfo {author} {\bibfnamefont {H.~C.}\
			\bibnamefont {Xu}}, \bibinfo {author} {\bibfnamefont {B.~P.}\ \bibnamefont
			{Xie}}, \bibinfo {author} {\bibfnamefont {J.~P.}\ \bibnamefont {Hu}},
		\bibinfo {author} {\bibfnamefont {T.}~\bibnamefont {Zhang}}, \ and\ \bibinfo
		{author} {\bibfnamefont {D.~L.}\ \bibnamefont {Feng}},\ }\href
	{http://dx.doi.org/10.1038/nphys3450} {\bibfield  {journal} {\bibinfo
			{journal} {Nat. Phys.}\ }\textbf {\bibinfo {volume} {11}},\ \bibinfo {pages}
		{946} (\bibinfo {year} {2015})}\BibitemShut {NoStop}%
	\bibitem [{\citenamefont {Liu}\ \emph {et~al.}(2018)\citenamefont {Liu},
		\citenamefont {Mao}, \citenamefont {Ding}, \citenamefont {Wu}, \citenamefont
		{Tang}, \citenamefont {Li}, \citenamefont {He}, \citenamefont {Li},
		\citenamefont {Song}, \citenamefont {Ma}, \citenamefont {Liu}, \citenamefont
		{Wang},\ and\ \citenamefont {Xue}}]{Liu2018_2}%
	\BibitemOpen
	\bibfield  {author} {\bibinfo {author} {\bibfnamefont {C.}~\bibnamefont
			{Liu}}, \bibinfo {author} {\bibfnamefont {J.}~\bibnamefont {Mao}}, \bibinfo
		{author} {\bibfnamefont {H.}~\bibnamefont {Ding}}, \bibinfo {author}
		{\bibfnamefont {R.}~\bibnamefont {Wu}}, \bibinfo {author} {\bibfnamefont
			{C.}~\bibnamefont {Tang}}, \bibinfo {author} {\bibfnamefont {F.}~\bibnamefont
			{Li}}, \bibinfo {author} {\bibfnamefont {K.}~\bibnamefont {He}}, \bibinfo
		{author} {\bibfnamefont {W.}~\bibnamefont {Li}}, \bibinfo {author}
		{\bibfnamefont {C.-L.}\ \bibnamefont {Song}}, \bibinfo {author}
		{\bibfnamefont {X.-C.}\ \bibnamefont {Ma}}, \bibinfo {author} {\bibfnamefont
			{Z.}~\bibnamefont {Liu}}, \bibinfo {author} {\bibfnamefont {L.}~\bibnamefont
			{Wang}}, \ and\ \bibinfo {author} {\bibfnamefont {Q.-K.}\ \bibnamefont
			{Xue}},\ }\href {\doibase 10.1103/PhysRevB.97.024502} {\bibfield  {journal}
		{\bibinfo  {journal} {Phys. Rev. B}\ }\textbf {\bibinfo {volume} {97}},\
		\bibinfo {pages} {024502} (\bibinfo {year} {2018})}\BibitemShut {NoStop}%
	\bibitem [{\citenamefont {Ge}\ \emph {et~al.}(2019)\citenamefont {Ge},
		\citenamefont {Yan}, \citenamefont {Zhang}, \citenamefont {Agterberg},
		\citenamefont {Weinert},\ and\ \citenamefont {Li}}]{Ge2019}%
	\BibitemOpen
	\bibfield  {author} {\bibinfo {author} {\bibfnamefont {Z.}~\bibnamefont
			{Ge}}, \bibinfo {author} {\bibfnamefont {C.}~\bibnamefont {Yan}}, \bibinfo
		{author} {\bibfnamefont {H.}~\bibnamefont {Zhang}}, \bibinfo {author}
		{\bibfnamefont {D.}~\bibnamefont {Agterberg}}, \bibinfo {author}
		{\bibfnamefont {M.}~\bibnamefont {Weinert}}, \ and\ \bibinfo {author}
		{\bibfnamefont {L.}~\bibnamefont {Li}},\ }\href {\doibase
		10.1021/acs.nanolett.9b00135} {\bibfield  {journal} {\bibinfo  {journal}
			{Nano Lett.}\ }\textbf {\bibinfo {volume} {19}},\ \bibinfo {pages} {2497}
		(\bibinfo {year} {2019})}\BibitemShut {NoStop}%
	\bibitem [{\citenamefont {Liu}\ \emph {et~al.}(2019{\natexlab{b}})\citenamefont
		{Liu}, \citenamefont {Wang}, \citenamefont {Gao}, \citenamefont {Liu},
		\citenamefont {Liu}, \citenamefont {Wang},\ and\ \citenamefont
		{Wang}}]{LiuPRL2019}%
	\BibitemOpen
	\bibfield  {author} {\bibinfo {author} {\bibfnamefont {C.}~\bibnamefont
			{Liu}}, \bibinfo {author} {\bibfnamefont {Z.}~\bibnamefont {Wang}}, \bibinfo
		{author} {\bibfnamefont {Y.}~\bibnamefont {Gao}}, \bibinfo {author}
		{\bibfnamefont {X.}~\bibnamefont {Liu}}, \bibinfo {author} {\bibfnamefont
			{Y.}~\bibnamefont {Liu}}, \bibinfo {author} {\bibfnamefont {Q.-H.}\
			\bibnamefont {Wang}}, \ and\ \bibinfo {author} {\bibfnamefont
			{J.}~\bibnamefont {Wang}},\ }\href {\doibase 10.1103/PhysRevLett.123.036801}
	{\bibfield  {journal} {\bibinfo  {journal} {Phys. Rev. Lett.}\ }\textbf
		{\bibinfo {volume} {123}},\ \bibinfo {pages} {036801} (\bibinfo {year}
		{2019}{\natexlab{b}})}\BibitemShut {NoStop}%
	\bibitem [{\citenamefont {Zhang}\ \emph {et~al.}(2020)\citenamefont {Zhang},
		\citenamefont {Ge}, \citenamefont {Weinert},\ and\ \citenamefont
		{Li}}]{Zhang2020}%
	\BibitemOpen
	\bibfield  {author} {\bibinfo {author} {\bibfnamefont {H.}~\bibnamefont
			{Zhang}}, \bibinfo {author} {\bibfnamefont {Z.}~\bibnamefont {Ge}}, \bibinfo
		{author} {\bibfnamefont {M.}~\bibnamefont {Weinert}}, \ and\ \bibinfo
		{author} {\bibfnamefont {L.}~\bibnamefont {Li}},\ }\href
	{https://doi.org/10.1038/s42005-020-0351-1} {\bibfield  {journal} {\bibinfo
			{journal} {Commun. Phys.}\ }\textbf {\bibinfo {volume} {3}},\ \bibinfo
		{pages} {75} (\bibinfo {year} {2020})}\BibitemShut {NoStop}%
	\bibitem [{\citenamefont {Kubo}(2007)}]{Kubo2007}%
	\BibitemOpen
	\bibfield  {author} {\bibinfo {author} {\bibfnamefont {K.}~\bibnamefont
			{Kubo}},\ }\href {\doibase 10.1103/PhysRevB.75.224509} {\bibfield  {journal}
		{\bibinfo  {journal} {Phys. Rev. B}\ }\textbf {\bibinfo {volume} {75}},\
		\bibinfo {pages} {224509} (\bibinfo {year} {2007})}\BibitemShut {NoStop}%
	\bibitem [{\citenamefont {Graser}\ \emph {et~al.}(2009)\citenamefont {Graser},
		\citenamefont {Maier}, \citenamefont {Hirschfeld},\ and\ \citenamefont
		{Scalapino}}]{Graser2009}%
	\BibitemOpen
	\bibfield  {author} {\bibinfo {author} {\bibfnamefont {S.}~\bibnamefont
			{Graser}}, \bibinfo {author} {\bibfnamefont {T.~A.}\ \bibnamefont {Maier}},
		\bibinfo {author} {\bibfnamefont {P.~J.}\ \bibnamefont {Hirschfeld}}, \ and\
		\bibinfo {author} {\bibfnamefont {D.~J.}\ \bibnamefont {Scalapino}},\ }\href
	{\doibase 10.1088/1367-2630/11/2/025016} {\bibfield  {journal} {\bibinfo
			{journal} {New J. Phys.}\ }\textbf {\bibinfo {volume} {11}},\ \bibinfo
		{pages} {025016} (\bibinfo {year} {2009})}\BibitemShut {NoStop}%
	\bibitem [{\citenamefont {Hao}\ and\ \citenamefont {Hu}(2014)}]{Hao2014}%
	\BibitemOpen
	\bibfield  {author} {\bibinfo {author} {\bibfnamefont {N.}~\bibnamefont
			{Hao}}\ and\ \bibinfo {author} {\bibfnamefont {J.}~\bibnamefont {Hu}},\
	}\href {\doibase 10.1103/PhysRevX.4.031053} {\bibfield  {journal} {\bibinfo
			{journal} {Phys. Rev. X}\ }\textbf {\bibinfo {volume} {4}},\ \bibinfo {pages}
		{031053} (\bibinfo {year} {2014})}\BibitemShut {NoStop}%
	\bibitem [{\citenamefont {Eschrig}\ and\ \citenamefont
		{Koepernik}(2009)}]{Eschrig2009}%
	\BibitemOpen
	\bibfield  {author} {\bibinfo {author} {\bibfnamefont {H.}~\bibnamefont
			{Eschrig}}\ and\ \bibinfo {author} {\bibfnamefont {K.}~\bibnamefont
			{Koepernik}},\ }\href {\doibase 10.1103/PhysRevB.80.104503} {\bibfield
		{journal} {\bibinfo  {journal} {Phys. Rev. B}\ }\textbf {\bibinfo {volume}
			{80}},\ \bibinfo {pages} {104503} (\bibinfo {year} {2009})}\BibitemShut
	{NoStop}%
	\bibitem [{\citenamefont {Lenck}\ \emph {et~al.}(1994)\citenamefont {Lenck},
		\citenamefont {Carbotte},\ and\ \citenamefont {Dynes}}]{Lenck1994}%
	\BibitemOpen
	\bibfield  {author} {\bibinfo {author} {\bibfnamefont {S.}~\bibnamefont
			{Lenck}}, \bibinfo {author} {\bibfnamefont {J.~P.}\ \bibnamefont {Carbotte}},
		\ and\ \bibinfo {author} {\bibfnamefont {R.~C.}\ \bibnamefont {Dynes}},\
	}\href {\doibase 10.1103/PhysRevB.49.9111} {\bibfield  {journal} {\bibinfo
			{journal} {Phys. Rev. B}\ }\textbf {\bibinfo {volume} {49}},\ \bibinfo
		{pages} {9111} (\bibinfo {year} {1994})}\BibitemShut {NoStop}%
	\bibitem [{\citenamefont {Migdal}(1958)}]{Migdal1958}%
	\BibitemOpen
	\bibfield  {author} {\bibinfo {author} {\bibfnamefont {A.}~\bibnamefont
			{Migdal}},\ }\href@noop {} {\bibfield  {journal} {\bibinfo  {journal} {Sov.
				Phys. JETP}\ }\textbf {\bibinfo {volume} {34}},\ \bibinfo {pages} {996}
		(\bibinfo {year} {1958})}\BibitemShut {NoStop}%
	\bibitem [{\citenamefont {Eliashberg}(1960)}]{Eliashberg1960}%
	\BibitemOpen
	\bibfield  {author} {\bibinfo {author} {\bibfnamefont {G.~M.}\ \bibnamefont
			{Eliashberg}},\ }\href@noop {} {\bibfield  {journal} {\bibinfo  {journal}
			{Sov. Phys. JETP}\ }\textbf {\bibinfo {volume} {11}},\ \bibinfo {pages} {696}
		(\bibinfo {year} {1960})}\BibitemShut {NoStop}%
	\bibitem [{Upp()}]{UppSC}%
	\BibitemOpen
	\href@noop {} {}\bibinfo {note} {The Uppsala Superconductivity (UppSC) code
		provides a package to self-consistently solve the anisotropic, multiband, and
		full-bandwidth Eliashberg equations for frequency-even and odd
		superconductivity mediated by phonons or spin fluctuations on the basis of
		\textit{ab initio} calculated input.}\BibitemShut {Stop}%
	\bibitem [{\citenamefont {Aperis}\ \emph {et~al.}(2015)\citenamefont {Aperis},
		\citenamefont {Maldonado},\ and\ \citenamefont {Oppeneer}}]{Aperis2015}%
	\BibitemOpen
	\bibfield  {author} {\bibinfo {author} {\bibfnamefont {A.}~\bibnamefont
			{Aperis}}, \bibinfo {author} {\bibfnamefont {P.}~\bibnamefont {Maldonado}}, \
		and\ \bibinfo {author} {\bibfnamefont {P.~M.}\ \bibnamefont {Oppeneer}},\
	}\href {\doibase 10.1103/PhysRevB.92.054516} {\bibfield  {journal} {\bibinfo
			{journal} {Phys. Rev. B}\ }\textbf {\bibinfo {volume} {92}},\ \bibinfo
		{pages} {054516} (\bibinfo {year} {2015})}\BibitemShut {NoStop}%
	\bibitem [{\citenamefont {Bekaert}\ \emph {et~al.}(2018)\citenamefont
		{Bekaert}, \citenamefont {Aperis}, \citenamefont {Partoens}, \citenamefont
		{Oppeneer},\ and\ \citenamefont {Milo\ifmmode \check{s}\else
			\v{s}\fi{}evi\ifmmode~\acute{c}\else \'{c}\fi{}}}]{Bekaert2018}%
	\BibitemOpen
	\bibfield  {author} {\bibinfo {author} {\bibfnamefont {J.}~\bibnamefont
			{Bekaert}}, \bibinfo {author} {\bibfnamefont {A.}~\bibnamefont {Aperis}},
		\bibinfo {author} {\bibfnamefont {B.}~\bibnamefont {Partoens}}, \bibinfo
		{author} {\bibfnamefont {P.~M.}\ \bibnamefont {Oppeneer}}, \ and\ \bibinfo
		{author} {\bibfnamefont {M.~V.}\ \bibnamefont {Milo\ifmmode \check{s}\else
				\v{s}\fi{}evi\ifmmode~\acute{c}\else \'{c}\fi{}}},\ }\href {\doibase
		10.1103/PhysRevB.97.014503} {\bibfield  {journal} {\bibinfo  {journal} {Phys.
				Rev. B}\ }\textbf {\bibinfo {volume} {97}},\ \bibinfo {pages} {014503}
		(\bibinfo {year} {2018})}\BibitemShut {NoStop}%
	\bibitem [{\citenamefont {Schrodi}\ \emph {et~al.}(2019)\citenamefont
		{Schrodi}, \citenamefont {Aperis},\ and\ \citenamefont
		{Oppeneer}}]{Schrodi2019}%
	\BibitemOpen
	\bibfield  {author} {\bibinfo {author} {\bibfnamefont {F.}~\bibnamefont
			{Schrodi}}, \bibinfo {author} {\bibfnamefont {A.}~\bibnamefont {Aperis}}, \
		and\ \bibinfo {author} {\bibfnamefont {P.~M.}\ \bibnamefont {Oppeneer}},\
	}\href {\doibase 10.1103/PhysRevB.99.184508} {\bibfield  {journal} {\bibinfo
			{journal} {Phys. Rev. B}\ }\textbf {\bibinfo {volume} {99}},\ \bibinfo
		{pages} {184508} (\bibinfo {year} {2019})}\BibitemShut {NoStop}%
	\bibitem [{\citenamefont {Schrodi}\ \emph
		{et~al.}(2020{\natexlab{b}})\citenamefont {Schrodi}, \citenamefont
		{Oppeneer},\ and\ \citenamefont {Aperis}}]{Schrodi2020_2}%
	\BibitemOpen
	\bibfield  {author} {\bibinfo {author} {\bibfnamefont {F.}~\bibnamefont
			{Schrodi}}, \bibinfo {author} {\bibfnamefont {P.~M.}\ \bibnamefont
			{Oppeneer}}, \ and\ \bibinfo {author} {\bibfnamefont {A.}~\bibnamefont
			{Aperis}},\ }\href {\doibase 10.1103/PhysRevB.102.024503} {\bibfield
		{journal} {\bibinfo  {journal} {Phys. Rev. B}\ }\textbf {\bibinfo {volume}
			{102}},\ \bibinfo {pages} {024503} (\bibinfo {year}
		{2020}{\natexlab{b}})}\BibitemShut {NoStop}%
	\bibitem [{\citenamefont {Zhang}\ \emph {et~al.}(2016)\citenamefont {Zhang},
		\citenamefont {Lee}, \citenamefont {Moore}, \citenamefont {Li}, \citenamefont
		{Yi}, \citenamefont {Hashimoto}, \citenamefont {Lu}, \citenamefont
		{Devereaux}, \citenamefont {Lee},\ and\ \citenamefont {Shen}}]{Zhang2016}%
	\BibitemOpen
	\bibfield  {author} {\bibinfo {author} {\bibfnamefont {Y.}~\bibnamefont
			{Zhang}}, \bibinfo {author} {\bibfnamefont {J.~J.}\ \bibnamefont {Lee}},
		\bibinfo {author} {\bibfnamefont {R.~G.}\ \bibnamefont {Moore}}, \bibinfo
		{author} {\bibfnamefont {W.}~\bibnamefont {Li}}, \bibinfo {author}
		{\bibfnamefont {M.}~\bibnamefont {Yi}}, \bibinfo {author} {\bibfnamefont
			{M.}~\bibnamefont {Hashimoto}}, \bibinfo {author} {\bibfnamefont {D.~H.}\
			\bibnamefont {Lu}}, \bibinfo {author} {\bibfnamefont {T.~P.}\ \bibnamefont
			{Devereaux}}, \bibinfo {author} {\bibfnamefont {D.-H.}\ \bibnamefont {Lee}},
		\ and\ \bibinfo {author} {\bibfnamefont {Z.-X.}\ \bibnamefont {Shen}},\
	}\href {\doibase 10.1103/PhysRevLett.117.117001} {\bibfield  {journal}
		{\bibinfo  {journal} {Phys. Rev. Lett.}\ }\textbf {\bibinfo {volume} {117}},\
		\bibinfo {pages} {117001} (\bibinfo {year} {2016})}\BibitemShut {NoStop}%
	\bibitem [{\citenamefont {Shigekawa}\ \emph {et~al.}(2018)\citenamefont
		{Shigekawa}, \citenamefont {Nakayama}, \citenamefont {Phan}, \citenamefont
		{Kuno}, \citenamefont {Sugawara}, \citenamefont {Takahashi},\ and\
		\citenamefont {Sato}}]{Shigekawa2018}%
	\BibitemOpen
	\bibfield  {author} {\bibinfo {author} {\bibfnamefont {K.}~\bibnamefont
			{Shigekawa}}, \bibinfo {author} {\bibfnamefont {K.}~\bibnamefont {Nakayama}},
		\bibinfo {author} {\bibfnamefont {G.~N.}\ \bibnamefont {Phan}}, \bibinfo
		{author} {\bibfnamefont {M.}~\bibnamefont {Kuno}}, \bibinfo {author}
		{\bibfnamefont {K.}~\bibnamefont {Sugawara}}, \bibinfo {author}
		{\bibfnamefont {T.}~\bibnamefont {Takahashi}}, \ and\ \bibinfo {author}
		{\bibfnamefont {T.}~\bibnamefont {Sato}},\ }\href {\doibase
		10.1088/1742-6596/1054/1/012019} {\bibfield  {journal} {\bibinfo  {journal}
			{J. Phys.: Conf. Series}\ }\textbf {\bibinfo {volume} {1054}},\ \bibinfo
		{pages} {012019} (\bibinfo {year} {2018})}\BibitemShut {NoStop}%
	\bibitem [{\citenamefont {Tian}\ \emph {et~al.}(2016)\citenamefont {Tian},
		\citenamefont {Zhang}, \citenamefont {Li}, \citenamefont {Wu}, \citenamefont
		{Wu}, \citenamefont {Sun}, \citenamefont {Zhou}, \citenamefont {Wang},
		\citenamefont {Ma}, \citenamefont {Xue},\ and\ \citenamefont
		{Zhao}}]{Tian2016}%
	\BibitemOpen
	\bibfield  {author} {\bibinfo {author} {\bibfnamefont {Y.~C.}\ \bibnamefont
			{Tian}}, \bibinfo {author} {\bibfnamefont {W.~H.}\ \bibnamefont {Zhang}},
		\bibinfo {author} {\bibfnamefont {F.~S.}\ \bibnamefont {Li}}, \bibinfo
		{author} {\bibfnamefont {Y.~L.}\ \bibnamefont {Wu}}, \bibinfo {author}
		{\bibfnamefont {Q.}~\bibnamefont {Wu}}, \bibinfo {author} {\bibfnamefont
			{F.}~\bibnamefont {Sun}}, \bibinfo {author} {\bibfnamefont {G.~Y.}\
			\bibnamefont {Zhou}}, \bibinfo {author} {\bibfnamefont {L.}~\bibnamefont
			{Wang}}, \bibinfo {author} {\bibfnamefont {X.}~\bibnamefont {Ma}}, \bibinfo
		{author} {\bibfnamefont {Q.-K.}\ \bibnamefont {Xue}}, \ and\ \bibinfo
		{author} {\bibfnamefont {J.}~\bibnamefont {Zhao}},\ }\href {\doibase
		10.1103/PhysRevLett.116.107001} {\bibfield  {journal} {\bibinfo  {journal}
			{Phys. Rev. Lett.}\ }\textbf {\bibinfo {volume} {116}},\ \bibinfo {pages}
		{107001} (\bibinfo {year} {2016})}\BibitemShut {NoStop}%
\end{thebibliography}


\vspace*{2cm}

%\appendix

{\bf SUPPLEMENTARY MATERIAL}\\

\setcounter{equation}{0}
%{\green here some reformating is needed}\\
%\section{Multichannel Eliashberg theory for spin fluctuations and electron phonon coupling}\label{appTheory}

%As stated in the main text, the systems Hamiltonian is given by
The Hamiltonian of the system is given by
\begin{align}
\hat{H} = \hat{H}_0 + \hat{H}_{\mathrm{ph}} + \hat{H}_{\mathrm{eph}} + \hat{H}_{\mathrm{int}} ,
\end{align}	
with kinetic term $\hat{H}_0=\sum_{\mathbf{k},p,q,\sigma} \xi_{\mathbf{k},p,q} \hat{c}^{\dagger}_{\mathbf{k},p,\sigma}\hat{c}^{~}_{\mathbf{k},q,\sigma}$. Here we use $\hat{c}^{\dagger}_{\mathbf{k},p,\sigma}$ and $\hat{c}^{~}_{\mathbf{k},p,\sigma}$ as electron creation and annihilation operators, where $\sigma$ denotes spin, $\mathbf{k}$ a wave vector and $p,q$ orbital indices. Further, $\xi_{\mathbf{k},p,q}$ denotes electron energies in orbital space. The lattice vibrations are represented by $\hat{H}_{\mathrm{ph}} =  \hbar\Omega \sum_{\mathbf{q}} \left(\hat{b}_{\mathbf{q}}^{\dagger}\hat{b}^{~}_{\mathbf{q}} + \frac{1}{2}\right)$, where $\hbar\Omega=81\,\mathrm{meV}$ is the characteristic Einstein phonon frequency {of the interfacial phonon} and $\hat{b}^{\dagger}_{\mathbf{q}}$ ($\hat{b}_{\mathbf{q}}$) creates (annihilates) a phonon with exchange momentum $\mathbf{q}$. The electron-phonon interaction (EPI) is given by $\hat{H}_{\mathrm{eph}}$ which describes an electron $\hat{c}^{~}_{\mathbf{k},q,\sigma}$ being scattered into the state $\hat{c}^{\dagger}_{\mathbf{k}',p,\sigma}$ via scattering matrix elements $g_{\mathbf{q},p,q}$ and phonon displacements $\hat{u}_{\mathbf{q}}=\hat{b}_{\mathbf{q}}^{\dagger}+\hat{b}_{-\mathbf{q}}$. In this work we set $\mathbf{q}=\mathbf{k}-\mathbf{k}'$.

Electron-electron interactions in the system are modeled via
\begin{align}
&\hat{H}_{\mathrm{int}} =  U \sum_{i,s}\hat{n}_{i,s,\uparrow}\hat{n}_{i,s,\downarrow} + \frac{V'}{2} \sum_{i,s,t\neq s} \hat{n}_{i,s}\hat{n}_{i,t} \nonumber\\
& - \frac{J}{2} \sum_{i,s,t\neq s} \hat{\vec{S}}_{i,s}\cdot\hat{\vec{S}}_{i,t}   + \frac{J'}{2} \sum_{i,s,t\neq s,\sigma} \hat{c}^{\dagger}_{i,s,\sigma} \hat{c}^{\dagger}_{i,s,\bar{\sigma}} \hat{c}_{i,t,\bar{\sigma}}\hat{c}_{i,t,\sigma} . \label{cs}
\end{align}	
In Eq.\,(\ref{cs}) we use index $i$ to describe the lattice site, hence $\hat{c}^{\dagger}_{i,s,\sigma}$ create an electron of orbital character $s$ with spin $\sigma$ at site $i$. We describe spin operators by $\hat{\vec{S}}_{i,s}$ as defined e.g.\ in Ref.\,\cite{Graser2009}. The occupation numbers are given by $\hat{n}_{i,s,\sigma}=\hat{c}^{\dagger}_{i,s,\sigma}\hat{c}^{~}_{i,s,\sigma}$ and $\hat{n}_{i,s}=\sum_{\sigma}\hat{c}^{\dagger}_{i,s,\sigma}\hat{c}^{~}_{i,s,\sigma}$. The first and third term scale with intraorbital onsite coupling $U$ and Hund's rule coupling $J$, respectively. For the interorbital onsite energy $V'$ and the pair hopping amplitude $J'$ we make the choice $V'=U-3J/4-J'$ and $J'=J/2$\,\cite{Kubo2007,Graser2009}.

A diagonalization of the hopping energies $\xi_{\mathbf{k},p,q}$ provides us with the electronic dispersion in band space $\xi_{\mathbf{k},n}$ and matrix elements $a_{\mathbf{k},n}^p$. These are used to calculate the bare susceptibilities
\begin{align}
&\mathrm{Im}\big(\big[\chi^0_{\mathbf{q}}(\omega)\big]_{st}^{pq}\big) = -\pi \sum_{n,n',\mathbf{k}'}  a_{\mathbf{k},n}^s a_{\mathbf{k},n}^{p,*} a_{\mathbf{k}+\mathbf{q},n'}^q a_{\mathbf{k}+\mathbf{q},n'}^{t,*} \nonumber \\
&~~~~~~~ \times \big[n_{\mathrm{F}}(\xi_{\mathbf{k},n}) - n_{\mathrm{F}}(\xi_{\mathbf{k}+\mathbf{q},n'})\big] \delta\big( \xi_{\mathbf{k}+\mathbf{q},n'} - \xi_{\mathbf{k},n} + \omega \big) ,\nonumber \\
&\mathrm{Re}\big(\big[\chi^0_{\mathbf{q}}(\omega)\big]_{st}^{pq}\big) = \frac{1}{\pi} \mathcal{P} \int_{-\infty}^{\infty} \frac{\mathrm{d}\omega'}{\omega'-\omega} \mathrm{Im}\big(\big[\chi^0_{\mathbf{q}}(\omega)\big]_{st}^{pq}\big)  ,
\end{align}
as function of frequency $\omega$, where $n_{\mathrm{F}}(.)$ is the Fermi-Dirac function. Within the Random Phase Approximation (RPA) the spin and charge susceptibilities are respectively calculated as
\begin{align}
\big[\chi^{\mathrm{S}}_{\mathbf{q}}(\omega)\big]_{st}^{pq} &= \big[\chi^0_{\mathbf{q}}(\omega)\big]_{st}^{pq} \nonumber\\
&+ \sum_{u,v,w,z} \big[\chi^{\mathrm{S}}_{\mathbf{q}}(\omega)\big]_{uv}^{pq} \big[U^{\mathrm{S}}\big]_{wz}^{uv} \big[\chi^0_{\mathbf{q}}(\omega)\big]_{st}^{wz},  \\
\big[\chi^{\mathrm{C}}_{\mathbf{q}}(\omega)\big]_{st}^{pq} &= \big[\chi^0_{\mathbf{q}}(\omega)\big]_{st}^{pq} \nonumber\\
&- \sum_{u,v,w,z} \big[\chi^{\mathrm{C}}_{\mathbf{q}}(\omega)\big]_{uv}^{pq} \big[U^{\mathrm{C}}\big]_{wz}^{uv} \big[\chi^0_{\mathbf{q}}(\omega)\big]_{st}^{wz}  ,
\end{align}
where we make use of the Stoner tensors
\begin{align}
& \big[U^{\mathrm{S}}\big]_{aa}^{aa}=U ~,~ \big[U^{\mathrm{S}}\big]_{bb}^{aa}=\frac{J}{2} ~,~ \big[U^{\mathrm{S}}\big]_{ab}^{ab}=\frac{J}{4}+V' ~,~   \nonumber \\
& \big[U^{\mathrm{S}}\big]_{ab}^{ba}=J' ~,~ \big[U^{\mathrm{C}}\big]_{aa}^{aa}=U ~,~ \big[U^{\mathrm{C}}\big]_{bb}^{aa}=2V' ~,~ \nonumber\\
& \big[U^{\mathrm{C}}\big]_{ab}^{ab}=\frac{3J}{4}-V' ~,~ \big[U^{\mathrm{C}}\big]_{ab}^{ba}=J'   ~.
\end{align}
The RPA susceptibilities can be used to find the available phase space for choosing $U$ and $J$, see Ref.\,\cite{Schrodi2020_3}.

From here we calculate the real-frequency dependent interaction kernels for spin and charge fluctuations via
\begin{align}
\big[V^{(+)}_{\mathbf{q}}(\omega)\big]_{st}^{pq}  =&  \big[\frac{3}{2} U^{\mathrm{S}} \chi^{\mathrm{S}}_{\mathbf{q}}(\omega) U^{\mathrm{S}} + \frac{1}{2}U^{\mathrm{C}} \chi^{\mathrm{C}}_{\mathbf{q}}(\omega)U^{\mathrm{C}} \big]_{ps}^{tq}  \label{vp}\\
\big[V^{(-)}_{\mathbf{q}}(\omega)\big]_{st}^{pq}  =&  \big[\frac{3}{2} U^{\mathrm{S}} \chi^{\mathrm{S}}_{\mathbf{q}}(\omega) U^{\mathrm{S}} + \frac{1}{2}U^{\mathrm{S}} \nonumber\\
&- \frac{1}{2}U^{\mathrm{C}} \chi^{\mathrm{C}}_{\mathbf{q}}(\omega)U^{\mathrm{C}} + \frac{1}{2}U^{\mathrm{C}} \big]_{ps}^{tq}  . \label{vm}
\end{align} 
The outcome of Eqs.\,(\ref{vp}) and (\ref{vm}) is used to compute the kernels in band space ($\mathbf{q}=\mathbf{k}-\mathbf{k}'$):
\begin{align}
&\big[V^{(\pm)}_{\mathbf{q}}(\omega)\big]_{n,n'} = \sum_{\mathbf{k},s,t,p,q} a^{t*}_{\mathbf{k},n} a^{s*}_{\mathbf{k},n} \big[V^{(\pm)}_{\mathbf{q}}(\omega)\big]_{st}^{pq} a^p_{\mathbf{k}',n'}a^q_{\mathbf{k}',n'}   .
\end{align}
Hereafter we obtain the Matsubara frequency-dependent interactions as
\begin{align}
& V^{(\pm)}_{\mathbf{q},l,n,n'} = \frac{1}{\pi} \mathcal{P} \int_{-\omega_{\mathrm{cut}}}^{\omega_{\mathrm{cut}}} \frac{\mathrm{d}\omega}{\omega-iq_m} \mathrm{Im}\big(\big[V^{(\pm)}_{\mathbf{q}}(\omega)\big]_{n,n'}\big)    \label{vpm}
\end{align}
with high-energy cutoff $\omega_{\mathrm{cut}}$ {which removes the high-energy parts of the excitation spectrum, especially the incoherent part that is irrelevant to superconductivity}\,\cite{Schrodi2020_3}. 

As stated before, the electron-phonon part of the interaction is given by an Einstein phonon spectrum, so we can write the kernel as
\begin{align}
V^{(\mathrm{eph})}_{\mathbf{q},l,n,n'} = |g_{\mathbf{q},n,n'}|^2 \frac{2\Omega}{\Omega^2 + q_l^2} ~, \label{epi}
\end{align}
with band-dependent scattering elements $g_{\mathbf{q},n,n'}$. In this work we make the approximation $g_{\mathbf{q},n,n'}=g_{\mathbf{q}}$, with $g_{\mathbf{q}}$ as given in the main text\,\cite{Aperis2018}.

The self-consistent Eliashberg equations that we solve in this work are derived in the standard way. Starting from the non-interacting electron Green's function in Nambu space $\big[\hat{G}^0_{\mathbf{k},m,n}\big]^{-1} = i\omega_m \hat{\rho}_0 - \xi_{\mathbf{k},n}\hat{\rho}_3$, we know that bosonic interactions lead to a Green's function $\hat{G}_{\mathbf{k},m,n}$, which obeys a Dyson equation $\hat{G}^{~}_{\mathbf{k},m,n} = \hat{G}^0_{\mathbf{k},m,n} + \hat{G}^0_{\mathbf{k},m,n}\hat{\Sigma}^{~}_{\mathbf{k},m,n} \hat{G}^{~}_{\mathbf{k},m,n}$. Here $\hat{\rho}_i$ are Pauli matrices and $\hat{\Sigma}_{\mathbf{k},m,n}$ represents the electron self-energy. By defining
\begin{align}
\hat{G}^{-1}_{\mathbf{k},m,n} = i\omega_m Z_{\mathbf{k},m,n}\hat{\rho}_0 - \big(\xi_{\mathbf{k},n} + \Gamma_{\mathbf{k},m,n}\big)\hat{\rho}_3 - \phi_{\mathbf{k},m,n}\hat{\rho}_1
\end{align}
the self-energy takes the form
\begin{align}
\!\! \hat{\Sigma}_{\mathbf{k},m,n} = i\omega_m \big(1 - Z_{\mathbf{k},m,n}\big)\hat{\rho}_0 + \Gamma_{\mathbf{k},m,n}\hat{\rho}_3 + \phi_{\mathbf{k},m,n}\hat{\rho}_1 . \label{sigmadef}
\end{align}
Further, we express $\hat{\Sigma}_{\mathbf{k},m,n}$ as sum over all first order scattering processes due to EPI, spin fluctuations (SFs), and charge fluctuations (CFs)\,\cite{Lenck1994}:
\begin{align} 
\hat{\Sigma}_{\mathbf{k},m,n} = T \sum_{\mathbf{k}',m',n'} V^{\mathrm{eph}}_{\mathbf{k}-\mathbf{k}',m-m',n,n'} \hat{\rho}_3 \hat{G}_{\mathbf{k}',m',n'} \hat{\rho}_3 \nonumber \\
+ T \sum_{\mathbf{k}',m',n'} P^{\mathrm{S}}_{\mathbf{k}-\mathbf{k}',m-m',n,n'} \hat{\rho}_0 \hat{G}_{\mathbf{k}',m',n'} \hat{\rho}_0 \nonumber \\
+ T \sum_{\mathbf{k}',m',n'} P^{\mathrm{C}}_{\mathbf{k}-\mathbf{k}',m-m',n,n'} \hat{\rho}_3 \hat{G}_{\mathbf{k}',m',n'} \hat{\rho}_3 .\label{sigma-SM}
\end{align} 

In Eq.\,(\ref{sigma-SM}) the variables $V^{\mathrm{eph}}_{\mathbf{k}-\mathbf{k}',m-m',n,n'}$, $P^{\mathrm{S}}_{\mathbf{k}-\mathbf{k}',m-m',n,n'}$ and $P^{\mathrm{C}}_{\mathbf{k}-\mathbf{k}',m-m',n,n'}$ model the interactions due to phonons, SFs, and CFs, respectively. The EPI is given by Eq.\,(\ref{epi}), while the spin and charge terms are related to Eq.\,(\ref{vpm}) as
\begin{align}
&P^{\mathrm{S}}_{\mathbf{k}-\mathbf{k}',m-m',n,n'} + P^{\mathrm{C}}_{\mathbf{k}-\mathbf{k}',m-m',n,n'} =  V^{(+)}_{\mathbf{q},l,n,n'} \\
&P^{\mathrm{S}}_{\mathbf{k}-\mathbf{k}',m-m',n,n'} - P^{\mathrm{C}}_{\mathbf{k}-\mathbf{k}',m-m',n,n'} =  V^{(-)}_{\mathbf{q},l,n,n'} .
\end{align}
We now project the electron self-energy expressions Eqs.\ (\ref{sigma}) and (\ref{sigmadef}) onto $\hat{\rho}_0$, $\hat{\rho}_3$ and $\hat{\rho}_1$, respectively, leading to the Eliashberg equations
\begin{align}
\!\! \!\! Z_{\mathbf{k},n,m} &= 1 - \frac{T}{\omega_m} \sum_{\mathbf{k}',m',n'} K^{(+)}_{\mathbf{k}-\mathbf{k}',m-m',n,n'} \frac{\omega_{m'}Z_{\mathbf{k}',n',m'}}{\Theta_{\mathbf{k}',n',m'}}, \\
\Gamma_{\mathbf{k},n,m} &= T \sum_{\mathbf{k}',m',n'} K^{(+)}_{\mathbf{k}-\mathbf{k}',m-m',n,n'} \frac{\xi_{\mathbf{k}',n'} + \Gamma_{\mathbf{k}',n',m'}}{\Theta_{\mathbf{k}',n',m'}} ,  \\
\phi_{\mathbf{k},n,m} &=  -T \sum_{\mathbf{k}',m',n'} K^{(-)}_{\mathbf{k}-\mathbf{k}',m-m',n,n'} \frac{\phi_{\mathbf{k}',n',m'}}{\Theta_{\mathbf{k}',n',m'}} , \\
\Theta_{\mathbf{k},n,m} &=  \big[i\omega_m Z_{\mathbf{k},n,m}^2\big]^2 - \big[\xi_{\mathbf{k},n} + \Gamma_{\mathbf{k},n,m}\big]^2 -  \phi^2_{\mathbf{k},n,m} ,
\end{align}
for the mass renormalization $Z_{\mathbf{k},m,n}$, the chemical potential shift $\Gamma_{\mathbf{k},m,n}$, and order parameter $\phi_{\mathbf{k},m,n}$\,\cite{Schrodi2020_3}. Via kernels
\begin{align}
K^{(\pm)}_{\mathbf{q},l,n,n'} = V^{(\mathrm{eph})}_{\mathbf{q},l,n,n'} \pm V^{(\pm)}_{\mathbf{q},l,n,n'}
\end{align}
we treat all three bosonic Cooper pair mediators self-consistently on the same footing. 
%{\blue Details on the evaluation of the Matsubara frequency sums are given in Ref.\  \cite{Schrodi2019}.}

\end{document}